\documentclass[iop, revtex4]{emulateapj}

\usepackage{natbib}
\usepackage[pdftex]{lscape}
\usepackage{amsfonts,amsmath,amssymb} 
\usepackage{color}
\usepackage[backref,breaklinks,colorlinks,citecolor=blue]{hyperref} 
\usepackage[all]{hypcap} 

\def \Msol{{\rm M}_{\odot}}
\def \Zsol{{\rm Z}_{\odot}}
\def \logm{\log(M/\Msol)}
\def \lya{Ly$\alpha$~}

\def \oabund{12+\log({\rm O/H})}

\def \hbeta{H$\beta$}
\def \halpha{H$\alpha$}

\def \oii{[O{\scriptsize\,II}]}
\def \oiii{[O{\scriptsize\,III}]}

\def \hei{He{\scriptsize\,I}}

\def \NII{[N{\scriptsize\,II}]}
\def \SII{[S{\scriptsize\,II}]}

\def \oiiired{[O{\scriptsize\,III}]$\lambda5007$}
\def \oiiiblue{[O{\scriptsize\,III}]$\lambda4960$}

\newcommand{\myrefsec}[1]{\hyperref[#1]{Section~\ref*{#1}}}
\newcommand{\myreffig}[3]{\hyperref[#1]{Figures~\ref*{#1}}, \hyperref[#2]{\ref*{#2}}, and \hyperref[#3]{\ref*{#3}} }

\def \c3645{$[3.6]~-~[4.5]$}
\def \cK36{$[{\rm K}]~-~[3.6]$}
\def \cHK{$[{\rm H}]~-~[{\rm K}]$}

\def \EWha{EW(H${\rm \alpha}$)}

\def \EWhaz0{{\rm EW(H$\alpha$)$_0$}}
\def \ebmv{${\rm E(B-V)}$}
\def \ebmvneb{${\rm E(B-V)}_{\rm neb}$}
\def \ebmvstel{${\rm E(B-V)}_{\rm stel}$}
\def \ebmvz0{${\rm E(B-V)}_0$}

\slugcomment{}

\shorttitle{Coherent study of emission lines and sSFR at $3 < z < 6$}
\shortauthors{A. L. Faisst, et al.}

\begin{document}


\title{A coherent study of emission lines from broad-band photometry: specific star-formation rates and \oiii/\hbeta~ratio at $3 < z < 6$}


\author{
A. L. Faisst\altaffilmark{1,2,$\dagger$},
P. Capak\altaffilmark{1,2},
B. C. Hsieh\altaffilmark{3},
C. Laigle\altaffilmark{4},
M. Salvato\altaffilmark{5},
L. Tasca\altaffilmark{6},
P. Cassata\altaffilmark{6},
I. Davidzon\altaffilmark{6,7},
O. Ilbert\altaffilmark{6},
O. Le F\`evre\altaffilmark{6},
D. Masters\altaffilmark{1,2},
H. J. McCracken\altaffilmark{4},
C. Steinhardt\altaffilmark{1,2},
J. D. Silverman\altaffilmark{8},
S. de Barros\altaffilmark{7},
G. Hasinger\altaffilmark{9},
N. Z. Scoville\altaffilmark{2},
}


\affil{$^{1}$Infrared Processing and Analysis Center, California Institute of Technology, Pasadena, CA 91125, USA}
\affil{$^{2}$Cahill Center for Astronomy and Astrophysics, California Institute of Technology, Pasadena, CA 91125, USA}
\affil{$^{3}$ Academia Sinica, Institute of Astronomy \& Astrophysics, P.O. Box 23-141, Taipei 10617, Taiwan}
\affil{$^4$Institut d'Astrophysique de Paris, CNRS \& UPMC, UMR 7095, 98 bis Boulevard Arago, 75014, Paris, France}
\affil{$^{5}$Max Planck Institut f\"ur Extraterrestrische Physik, Giessenbachstrasse 1, D-85748, Garching bei M\"unchen, Germany}
\affil{$^{6}$Aix Marseille Universit\'e, CNRS, LAM (Laboratoire d'Astrophysique de Marseille) UMR 7326, 13388, Marseille, France}
\affil{$^{7}$INAF  -  Osservatorio  Astronomico  di  Bologna,  via  Ranzani  1,  I-40127, Bologna, Italy}
\affil{$^{8}$Kavli Institute for the Physics and Mathematics of the Universe (WPI), The University of Tokyo Institutes for Advanced Study, The University of Tokyo, Kashiwa, Chiba 277-8583, Japan}
\affil{$^{9}$ Institute for Astronomy, 2680 Woodlawn Dr., University of Hawaii, Honolulu, HI 96822, USA}



\affil{\textit{submitted to ApJ - \today}}

\email{afaisst@ipac.caltech.edu}

\altaffiltext{$\dagger$}{Twitter: @astrofaisst}


\begin{abstract}
	We measure the \halpha~and \oiii~emission line properties as well as specific star-formation rates (sSFR) of spectroscopically confirmed $3 < z < 6$ galaxies in COSMOS from their observed colors vs. redshift evolution. Our model describes consistently the \textit{ensemble of galaxies} including intrinsic properties (age, metallicity, star-formation history), dust-attenuation, and optical emission lines. We \textit{forward-model} the measured \halpha~equivalent-widths (EW) to obtain the sSFR out to $z\sim6$ without stellar mass fitting.
	We find a strongly increasing rest-frame \halpha~EW that is flattening off above $z\sim2.5$ with average EWs of $300-600\,{\rm \AA}$ at $z\sim6$. The sSFR is increasing proportional to $(1+z)^{2.4}$ at $z<2.2$ and $(1+z)^{1.5}$ at higher redshifts, indicative of a fast mass build-up in high-z galaxies within $e-$folding times of $100-200\,{\rm Myr}$ at $z\sim6$. The redshift evolution at $z>3$ cannot be fully explained in a picture of cold accretion driven growth.
	We find a progressively increasing \oiiired/\hbeta~ratio out to $z\sim6$, consistent with the ratios in local galaxies selected by increasing \halpha~EW (i.e., sSFR). This demonstrates the potential of using ``local high-z analogs'' to investigate the spectroscopic properties and relations of galaxies in the re-ionization epoch.
	
\end{abstract}





\section{Introduction}

With current broad-band near- to mid-infrared (IR) filters on ground- and space-based telescopes we are able to select galaxy samples in the very early epochs of the universe. However, the study of their physical properties~--~essential to refine our understanding of the formation and evolution of galaxies -- is hampered by several technical problems.

	In the recent years, a lot of progress has been made in understanding galaxy formation in the early universe before the peak of the cosmic star-formation density at $z\sim2-3$. In particular, several new avenues have been opened by large spectroscopic and photometric campaigns to explore the near- to mid-IR wavelength range on large parts of the sky.
	From these, it became clear that galaxies live on a so called ``main-sequence'' connecting their stellar mass with their star-formation rate (SFR) out to redshifts as high as $z\sim5$ \citep[][]{STEINHARDT14,SPEAGLE14,TASCA15}. Also, it is suggested that galaxies grow very rapidly in the early universe due to high gas fractions and/or star-formation efficiencies \citep[e.g.,][]{SCOVILLE15,SILVERMAN15}. Going in hand with the former, a marginal flattening of the relation between metallicity and stellar mass is expected for young galaxies at $z\sim5$ \citep[see][]{FAISST15}.
	These recent observations have triggered questions that have yet to be answered. For example, galaxies have been found that are more massive than expected from hierarchical assembly of dark-matter haloes \citep[e.g.,][]{STEINHARDT15}.
	The formation of such massive galaxies at high redshifts requires their fast growth and therefore an increase in the specific SFR (${\rm sSFR}={\rm SFR}/M$, a measure for the rate of mass build-up in galaxies) at $z>3$ \citep[e.g.,][]{WEINMANN11}. This increase is also predicted in the picture of accretion dominated galaxy growth \citep[e.g.,][]{DEKEL09,TACCHELLA13} and recent hydrodynamical simulations \citep[e.g.,][]{DAVE11,SPARRE15}. Some studies observe these predictions \citep[][]{STARK13,BARROS14,JIANG15}, while other find a considerable flattening of the sSFR at $z>3$ \citep{GONZALEZ14,TASCA15,MARMOL15}.	
	Finally, relations based on local galaxies, e.g., the relations between metallicity and strong emission lines, may not be applicable anymore at higher redshifts due to the change of internal physical processes in such galaxies such as ionization or the abundance of \NII~ \citep[e.g.,][]{MASTERS14,STEIDEL14,SHAPLEY15,COWIE15}.
	
	Stellar mass and SFRs as well as metallicity and ionization parameter are the most important basic physical quantities on which the above results are based on and the above questions depend on. While these can be measured reliably at low redshifts by a good multi-wavelength coverage in imaging and spectroscopy, there are several caveats at higher redshifts.
	First, SFRs have to be measured in the UV, as reliable estimators such as the \halpha~emission line are out of spectral coverage. The UV is highly sensitive to dust attenuation \citep[e.g.,][]{BOUWENS12a}, which shows a large diversity in high redshift galaxies \citep[see][]{CAPAK15}.
	Second, deep observed mid-IR imaging data are necessary to probe the old stellar populations in galaxies at $z>4$ and therefore allow a reliable measurement of stellar masses.
	Third, mid-IR filters at these redshifts are contaminated by the (unknown) contribution of strong emission lines, which boost the masses significantly \citep[e.g.,][]{SCHAERER09,STARK13,BARROS14}.
	Finally, the conversion from the observed data to these quantities (i.e., stellar mass and SFR) depends on theoretical models of the intrinsic properties of galaxies such as their age, metallicity, and star-formation history (SFH), all of which are not known a priori for individual galaxies at high redshifts.
	
	
	In this paper, we develop a model insensitive way to measure the sSFR and the emission line strength at $3 < z < 6$ from primary observables. Furthermore, we demonstrate the potential of using local high-z analogs to probe the spectral properties of high redshift galaxies up to $z=6$.
	In particular, we use the redshift evolution of the galaxy population averaged observed near- to mid-IR color to measure the \halpha~equivalent width (EW) from which we directly derive the sSFR$(z)$. The measurement of the EW has two parts, namely the measurement of the observed flux/color and the estimation of the underlying continuum between $4500\,{\rm \AA}$ and $6500\,{\rm \AA}$ underneath optical emission lines.
	The latter we \textit{forward model} by assuming intrinsic properties of the galaxies (dust attenuation, metallicity, stellar population age, star formation history) and we show that the resulting continuum is very insensitive to the choice of these parameters in the above specified wavelength range.
	The determination of the \halpha~equivalent width (EW) from observed galaxy colors has been used in the past \citep[][]{SHIM11,STARK13,SMIT15a,SMIT15b,RASAPPU15,MARMOL15}, however, mostly at discrete redshift bins and for small sample sizes. We perform here a consistent analysis across a large redshift range with a much larger sample of spectroscopically confirmed galaxies than in previous studies.
	Our large sample allows us to model the \textit{ensemble of galaxies} instead of considering single galaxies. This enables us to marginalize over the (poorly known) intrinsic properties of the galaxies when modeling the continuum below the optical emission lines. It also gives us a convenient way to describe the scatter (systematic and physical) of the ensemble's properties, which we can propagate through our model and investigate its effect on our results.
	
	
	The plan for this paper is as follows.
	In \myrefsec{sec:sample} we describe the sample of spectroscopically confirmed galaxies that is used for this analysis. 
	In \myrefsec{sec:model} we describe the modeling of the observed color vs. redshift relation including the contribution of intrinsic parameters (age, metallicity, SFH), dust attenuation, and optical emission lines.
	In the results section (\myrefsec{sec:results}), we derive the redshift evolution of the \halpha~EW, the sSFR$(z)$, as well as the \oiii/\halpha~ratio out to $z\sim6$.
	These results are discussed in \myrefsec{sec:discussion} and summarized in \myrefsec{sec:end}.
	
	Throughout this work we adopt a flat cosmology with $\Omega_{\Lambda,0}~=~0.7$, $\Omega_{m,0}~=~0.3$, and $h~=~0.7$. Magnitudes are given in the AB system \citep{OKE83} and all masses are scaled to a \citet{CHABRIER03} initial mass function (IMF).

\section{Data \& Sample selection}\label{sec:sample}

\subsection{Data}

	In this work, we use the two square degrees of the \textit{Cosmic Evolution Survey} \citep[COSMOS,][]{SCOVILLE07} field, which are observed by a wealth of instruments in imaging as well as spectroscopy across a broad range of wavelengths.
	We make use of the following data sets.
	
	\begin{enumerate}
		\item The COSMOS spectroscopy catalog, which contains more than $6000$ high-quality spectra at $1~<~z~<~6$ (M. Salvato, private communication).
		
		\item The VIMOS Ultra Deep Survey (VUDS) spectroscopy catalog, containing galaxy spectra at $2 < z < 6$ \citep[]{LEFEVRE14}.
		
		\item The \textit{COSMOS2015} photometric catalog including photometry from the UV to mid-IR as well as photometric redshifts and stellar masses \citep[][]{LAIGLE15}.
		
		
	\end{enumerate}

	The COSMOS spectroscopic master catalog available to the COSMOS collaboration is a compilation of all spectroscopic observations up to $z\sim6$ that are carried out on the COSMOS field. The galaxy sample is selected in different ways (color, photometric redshift, Lyman Break technique) and observed by several different instruments depending on the redshift (VIMOS, FORS2, FMOS, MOIRCS, DEIMOS, MOSFIRE). The different selection techniques lead to a large coverage of physical properties of the galaxies, thus this sample represents well the population of star-forming galaxies at these redshifts. For more information, we refer to the official COSMOS web-page\footnote{\url{http://cosmos.ipac.caltech.edu}}.
	
	The VUDS spectroscopy catalog contains galaxies selected by photometric redshifts with a flux limit of $i_{{\rm AB}}=25$. The spectra are obtained with the VIMOS spectrograph on the ESO Very Large Telescope \citep[][]{LEFEVRE03b}. For more information, we refer to \citet{LEFEVRE15}.
	
	The \textit{COSMOS2015} photometric catalog contains the photometry of the extracted galaxies on COSMOS measured from the UV to the mid-IR images. The source extraction is based on a $\chi^2$ image determined from the Subaru $z-$band and the COSMOS/UltraVISTA $YJHK$ bands \citep[see][]{CAPAK07,MCCRACKEN12,ILBERT13}. Part of this catalog is the mid-IR data at $3.6\,{\rm \mu m}$ and $4.5\,{\rm \mu m}$ down to $\sim 25.5~{\rm mag}$ ($3\,\sigma$ in 3$\arcsec$ diameter; as of October 2015) from the \textit{Spitzer Large Area Survey with Hyper-Suprime-Cam} \citep[SPLASH,][]{STEINHARDT14}\footnote{\url{http://splash.caltech.edu}}. These sources are extracted using the segmentation map of the \textit{COSMOS2015} catalog and an improved version of IRACLEAN \citep{HSIEH12} in order to overcome the source confusion (blending). 
	
	We subsequently match the photometry catalog with the spectroscopy catalogs within $1\arcsec$ radius in order to recover the photometry for our spectroscopically confirmed galaxies. More than $97\%$ of the galaxies are matched within a radius of $0.3\arcsec$.
	
	
	

\begin{figure}
\centering
\includegraphics[width=1.1\columnwidth, angle=0]{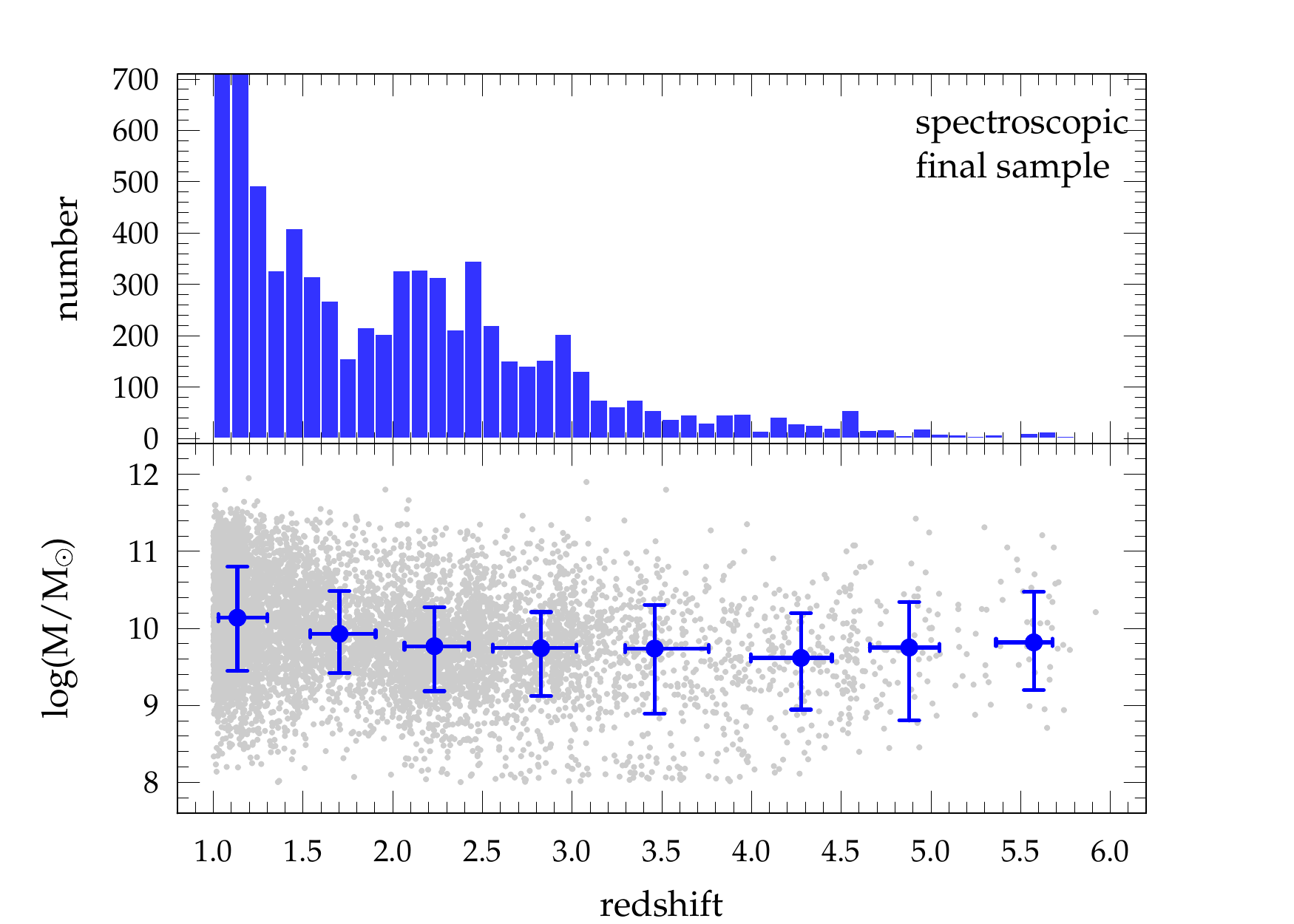}
\caption{Properties of our final sample (galaxies clear of contamination within $2\arcsec$). \textbf{Top:} Redshift distribution of our final spectroscopic sample.
\textbf{Bottom:} Stellar mass distribution as a function of redshift for our final spectroscopic sample of galaxies \citep[see][]{LAIGLE15}. The gray points show individual galaxies and the blue symbols show the mean $\logm$ in redshift bins with $68\%$ percentiles scatter in redshift and mass shown by the error bars.
\label{fig:mass}}
\end{figure}
	
\begin{figure*}
\centering
\includegraphics[width=2.1\columnwidth, angle=0]{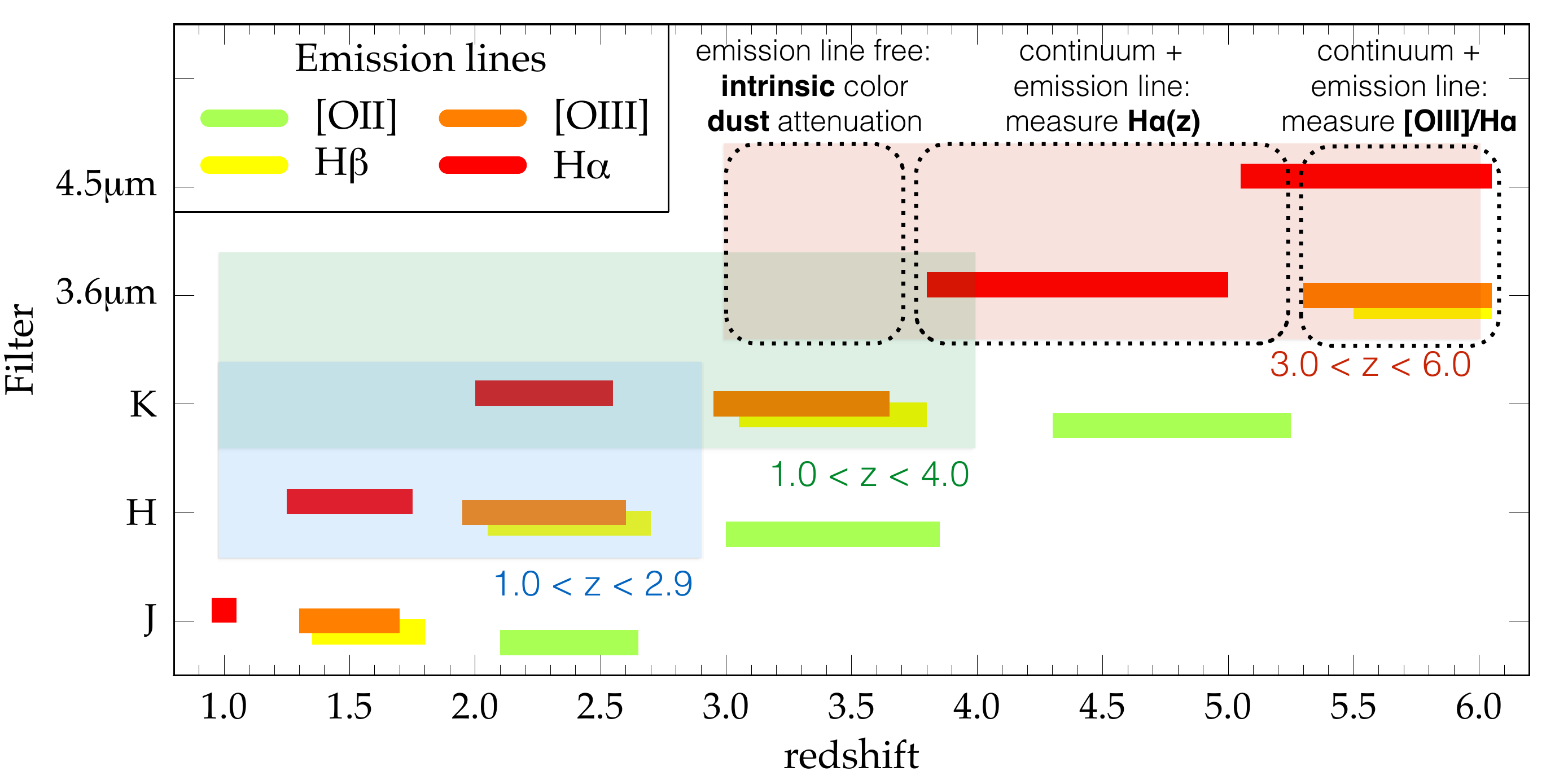}
\caption{Contribution of different optical emission lines (\oii, green; \oiii, orange; \halpha, red; \hbeta, yellow) to broad-band filters as a function of redshift. 
The blue, green, and red bands show the three redshift windows (redshifts indicated) to which we apply our emission line $+$ dust model to estimate the \halpha~EW as well as the \oiii/\halpha~line ratio. The dotted boxes show, for the $3 < z < 6$ window as textbook case, how redshift ranges are used to reveal the intrinsic color $+$ dust attenuation (used to anchor our model fit), the \EWha~vs. redshift evolution, and the ratio \oiii/\halpha.
\label{fig:emlins}}
\end{figure*}
	
	\subsection{Galaxy selection, redshift and stellar mass distribution}
	
	For the purpose of this work, we apply a very stringent cut to our sample in terms of both spectroscopy as well as photometry.
	We only include reliable spectroscopic redshifts in our sample ($>80\%$ probability of correct spectroscopic redshift) at $1 < z < 6$ and remove spectroscopically confirmed AGNs based on their broad optical emission. We will use the redshift range $1 < z < 3$ to verify our method by direct comparison of our results to spectroscopic emission line measurements.
	The measurement of colors strongly depends on source confusion, which commonly is taken into account during the extraction of the galaxy photometry. For the purpose of this work, we add an additional security and remove potentially blended sources in the near- to mid-IR by directly checking their number of neighboring galaxies. For this to end, we use the high-resolution F814W ($I-$band) images from the Hubble Space Telescope's Advanced Camera for Surveys (HST/ACS), as well as the COSMOS/UltraVISTA optical/near-IR selected catalog. We extract the number of companions within a certain aperture size for each galaxy in our sample. Given the PSF aperture size of $\sim2\arcsec-3\arcsec$ in the mid-IR, we remove all the galaxies with companions closer than $2\arcsec$ and $3\arcsec$ in radius, respectively.
	
	Our final sample at $1 < z < 6$ with no contamination of neighboring galaxies within a radius of $2\arcsec$ ($3\arcsec$) contains more than $4,000$ ($1,500$) galaxies in total. In the following analysis we use the galaxy sample clear of neighbors within $2\arcsec$. The results do not significantly change if we use the more restricted sample of galaxies without contamination within $3\arcsec$ radius (although the uncertainties are larger because of the strongly decreased number of galaxies). In particular, in the redshift range $3 < z < 6$, we use $530$ spectroscopically confirmed galaxies.
	\autoref{fig:mass} shows the stellar mass and redshift distribution of our final sample of galaxies (without contaminants within $2\arcsec$). The stellar masses are measured on the \textit{COSMOS2015} photometry using the SED fitting routine \texttt{Le Phare}\footnote{See \citet{ILBERT06} and \url{http://www.cfht.hawaii.edu/~arnouts/LEPHARE/}} including the fitting of emission lines set proportional to the UV flux \citep[][]{LAIGLE15}. We expect our galaxies to have a median stellar mass of $\logm \sim 9.8$ at $z>3$.
	\textit{We stress that the goal of this paper is to access the sSFR from primary measurements (galaxy color) and we therefore do not use these stellar masses in the following. They only serve to visualize the expected mass range of our sample of galaxies and the comparison to other studies}.

\section{Emission line strengths from observed colors}\label{sec:model}

	In this section, we describe in detail our model including intrinsic galaxy properties (age, metallicity, SFH), dust attenuation, and optical emission lines. From this we derive model galaxy colors as a function of redshift, which are compared to the observed colors vs. redshift evolution in our galaxy sample. This allows us, using a minimization algorithm, to solve for the spectral properties of the ensemble of these galaxies in specific redshift windows, which are detailed in the following.

\subsection{Redshift windows and colors}\label{sec:zwindows}
	
	The idea of this paper is to constrain the emission line properties of galaxies from their observed colors. Emission lines contribute to different broad-band filters for galaxies at different redshifts. This produces ''wiggles`` in the observed color-redshift evolution with respect to what is expected from a continuum without nebular lines.
	However, the observed color of a galaxy is not only affected by emission lines, but also by its intrinsic properties (age, metallicity, SFH) and dust attenuation. These change with redshift and thus are degenerate with the effects of emission lines. In a later section we will discuss how much these various properties affect the observed color of a galaxy. 
	In order to separate the effect of emission lines, we have to calibrate our model in redshift ranges in which the continuum flux in broad-band filters is free of emission lines and thus reveals the intrinsic color and dust attenuation.
		
	\autoref{fig:emlins} shows the location of strong emission lines (\halpha, \hbeta, \oii, and \oiii) in different near- and mid-IR broad-band filters as a function of redshift. There are several different redshift ranges (labeled for the case $3 < z < 6$ and mid-IR colors):
	
	\begin{itemize}
	\item Redshift ranges free of emission lines that reveal the intrinsic color and dust attenuation of the galaxies and thus anchor our model (box labeled with ``intrinsic/dust'').
	\item Redshift ranges where the observed color includes the \halpha~emission line and thus allows us to measure \EWha~(box labeled with ``\halpha'').
	\item Redshift ranges that allow the measurement of the the ratio \oiii/\halpha~(box labeled with ``\oiii/\halpha'').
	\end{itemize}
	
	These different redshift ranges exist for different observed colors and can be bundled in larger redshift windows. For the purpose of this work, we choose three different redshift windows, each with a corresponding observed color.
	
	\begin{itemize}
	\item[(A)] $1.0 < z < 2.9$ in observed \cHK~color,
	\item[(B)] $1.0 < z < 4.0$ in observed \cK36~color, and
	\item[(C)] $3.0 < z < 6.0$ in observed \c3645~color.
	\end{itemize}
	
	Each of the three redshift windows is designed to have a redshift range free of emission lines to anchor the model to the intrinsic color. Furthermore, this choice allows us to consistently model \EWha~across the redshift range $1 < z < 6$ and the \halpha/\oiii~ratio at $z\sim2.2$, $z\sim3.3$, and $z\sim5.5$. 
	The redshift windows (A) and (B) are used to verify our method by comparing our results to spectroscopic measurements.
	Given the strong dependence of the $4000~{\rm \AA}$ Balmer break on stellar population properties, we do not model the \oii~emission line here.
	Fortunately, the wavelength part red-ward of the \oii~emission is relatively insensitive to the intrinsic properties of the stellar population as we will discuss later.

\subsection{Modeling the mean observed color as a function of redshift}

	The observed color of a galaxy is affected threefold by its properties: \textit{(i)} By its intrinsic color (age, metallicity, SFH), \textit{(ii)} by the dust attenuation, and, \textit{(iii)} by emission lines.
	
	In the following sections, we build up a model for the observed color in the different redshift windows from these three contributions. From its comparison to the true observed colors, we can then compute the average emission line properties of our galaxies.
	It is important to note that the intrinsic galaxy properties as well as dust are solely used to represent the continuum under the \hbeta, \oiii, and \halpha~lines, i.e., red-ward of the 4000~{\rm \AA} break. In particular, the fitting of our model continuum to the observed continuum in line-free wavelength regions using the contribution from dust as a ``knob'' (\myrefsec{sec:emlindust}) smooths out possible variations in the intrinsic galaxy properties that are missed elsewhere in our model. This is the main advantage of our forward-modeling technique and allows the robust estimation of the emission line properties as we show in the following.
	
	
\subsubsection{The intrinsic color (age, metallicity, SFH)}\label{sec:intrinsic}

	For describing the intrinsic color of a galaxy population, we have to assume a metallicity, stellar population age, and SFH.
	These are unknown a priori, therefore we set up a grid that brackets reasonable choices of these parameters. Our forward-modeling technique then allows us to investigate the effects on the resulting observed color and the results derived in this work.
	As a basis we use the composite stellar population library from the \citet{BRUZUAL03} with a Chabrier IMF and create SEDs with different SFHs, metallicities, and ages using \texttt{GALAXEV}\footnote{See \citet{BRUZUAL03} and \url{http://www.bruzual.org/}}.
	
	Galaxies up to $z\sim4-5$ show a tight relation between SFR and stellar mass, which leads to an exponentially increasing SFH for the average population of galaxies \citep{NOESKE07,DADDI07,SPEAGLE14,STEINHARDT14b,SMIT15a}. We bracket possible histories by a constant SFH, a delayed exponentially decreasing SFH (${\rm SFR} \propto t/\tau_p^2\times\exp(-t/\tau_p)$) with a peak at $\tau_p=1\,{\rm Gyr}$, and an exponentially increasing SFH (${\rm SFR} \propto \exp(t/\tau)$) with an $e$-folding time of $\tau=500\,{\rm Myr}$.
	Furthermore, galaxies at high redshift show a considerably lower metallicity content \citep[e.g.,][]{ERB06,MAIOLINO08,MANNUCCI09,FAISST15}. We therefore bracket the range in metallicity between $Z=0.004$ (1/5$^{{\rm th}}$ of solar) and $Z=0.02$ (solar). However, because of the minor effect of metallicity on the continuum, we keep it constant with redshift. 
	Finally, we assume the age of the galaxy to be the time since the estimated start of re-ionization at $z=11$ \citep[e.g.,][]{PLANCK15}. Similar parameterizations of the galaxy's age as a function of redshift (e.g., half of the Hubble time) do not change our results.

\begin{figure}
\centering
\includegraphics[width=1.15\columnwidth, angle=0]{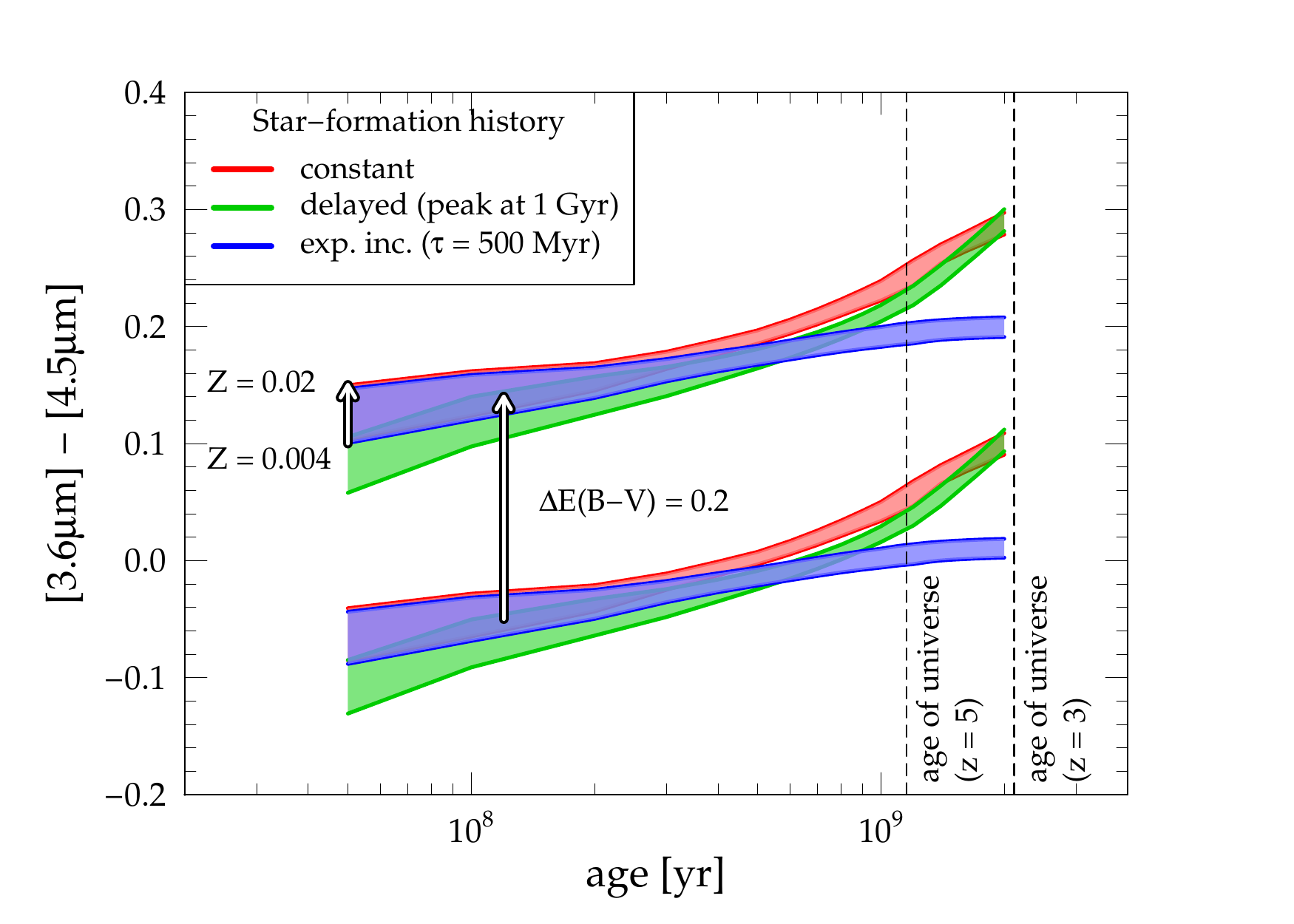}
\caption{Effect of intrinsic properties (metallicity, age, SFH) and dust on the observed \c3645~color in the case of a $z=5$ galaxy. Metallicity plays only a minor role in setting the color of a galaxy (see arrow from $1/5^{{\rm th}}$ of solar to solar). The colored bands show three different SFHs (constant, delayed, and exponentially increasing). Clearly, dust attenuation (ranging between \ebmv$=0.1-0.4\,{\rm mag}$ in our sample, depending on redshift) is the dominant contributor to color, followed by the SFH for galaxies older than $\sim1~{\rm Gyr}$ (or $z<5$). Importantly, the color is insensitive to reasonable SFHs and stellar population ages for young ($<1~{\rm Gyr}$) galaxies at high redshifts.
\label{fig:intrinsiccolor}}
\end{figure}
	
\capstartfalse
\begin{deluxetable*}{cc  cc |  ccc  cc | ccc}
\tabletypesize{\scriptsize}
\tablecaption{Summary of observational data and best-fit models.\label{tab:fitresults}}
\tablewidth{0pt}
\tablehead{
\multicolumn{4}{c}{Data and observations} & \multicolumn{5}{c}{Model fit$^\dagger$} & \multicolumn{3}{c}{Input properties\vspace{0cm}}   \\
\multicolumn{4}{c}{ } & \multicolumn{5}{c}{-----------------------------------------------------------------} & \multicolumn{3}{c}{ } \\
\multicolumn{4}{c}{ (spectroscopic)} & \multicolumn{3}{c}{Emission lines\vspace{0cm}} & \multicolumn{2}{c}{Dust$^b$\vspace{0cm}} & \multicolumn{3}{c}{  } \\
\multicolumn{4}{c}{------------------------------------------------------------} & \multicolumn{3}{c}{------------------------------------} & \multicolumn{2}{c}{---------------------------} & \multicolumn{3}{c}{--------------------------------------} \\
\colhead{redshift} & \colhead{color} & \colhead{ $\#<2\arcsec$ } & \colhead{ $\#<3\arcsec$ } & \colhead{\EWhaz0$^{a}$} & \colhead{$\alpha$} & \colhead{$\xi$} & \colhead{\ebmvz0} & \colhead{$z_{{\rm d,0}}$} & \colhead{Z [$\Zsol$]} & \colhead{Age [yr]} & \colhead{SFH}
}
\startdata\\[0.1cm]
$1.0 < z < 2.9$ & \cHK & 3571 & 1671 & 13.4 & 2.01 & 1.1 & 0.45 & 0.9 & 0.020 & $T(z)-T(11)$ & constant \\[0.2cm]
~ & ~ & ~ & ~ & 5.5 & 2.96 & 0.9 & 0.90 & 0.85 & 0.004 & $T(z)-T(11)$ & constant \\[0.2cm]
~ & ~ & ~ & ~ & 78.1 & 0.32 & 1.4 & 0.45 & 2.10 & 0.020 & $T(z)-T(11)$ & exp. inc \\[0.2cm]
~ & ~ & ~ & ~ & 57.2 & 0.72 & 1.0 & 0.85 & 1.40 & 0.04 & $T(z)-T(11)$ & exp. inc \\[0.2cm]

$1.0 < z < 4.0$ & \cK36 & 3863 & 1811 & 15.8 & 1.92 & 0.8 & 0.70 & 1.10 & 0.020 & $T(z)-T(11)$ & constant \\[0.2cm]
~ & ~ & ~ & ~ & 32.8 & 1.35 & 0.6 & 0.80 & 1.50 & 0.004 & $T(z)-T(11)$ & constant \\[0.2cm]
~ & ~ & ~ & ~ & 10.0 & 2.34 & 1.1 & 0.70 & 1.60 & 0.020 & $T(z)-T(11)$ & exp. inc \\[0.2cm]
~ & ~ & ~ & ~ & 21.0 & 1.71 & 0.7 & 0.90 & 1.80 & 0.004 & $T(z)-T(11)$ & exp. inc \\[0.2cm]

$3.0 < z < 6.0$ & \c3645 & 530 & 257 & 9.9 & 2.01 & 0.8$^{c}$ & 0.90 & 1.25 & 0.020 & $T(z)-T(11)$ & constant \\[0.2cm]
~ & ~ & ~ & ~ & 13.6 & 1.83 & 0.8$^{c}$ & 1.10 & 1.70 & 0.004 & $T(z)-T(11)$ & constant \\[0.2cm]
~ & ~ & ~ & ~ & 10.6 & 1.98 & 0.9$^{c}$ & 0.80 & 1.60 & 0.020 & $T(z)-T(11)$ & exp. inc. \\[0.2cm]
~ & ~ & ~ & ~ & 8.5 & 2.10 & 0.8$^{c}$ & 1.20 & 1.80 & 0.004 & $T(z)-T(11)$ & exp. inc. \\

\enddata
\tablenotetext{$\dagger$}{The errors in the resulting \EWha~are estimated by a Monte-Carlo simulation to be $\sim30\%$. The errors in $\xi=($\oiii/\halpha$)^{-1}$~are similarly estimated to be on the order of $70\%$.}
\tablenotetext{a}{In angstroms and rest-frame.}
\tablenotetext{b}{Strictly speaking, these parameters do not only include dust but also changes in the intrinsic SED that are not included elsewhere in our model.}
\tablenotetext{c}{These are best fit values, however, more uncertain than at $1.0 < z < 2.9$ because only part of the wavelength range including \oiii~and~\halpha~emission lines is covered by our data. The actual scatter in these measurements will be discussed in more detail in \myrefsec{sec:oiiiha}.}
\end{deluxetable*}
\capstarttrue

\begin{figure*}
\centering
\includegraphics[width=1.7\columnwidth, angle=0]{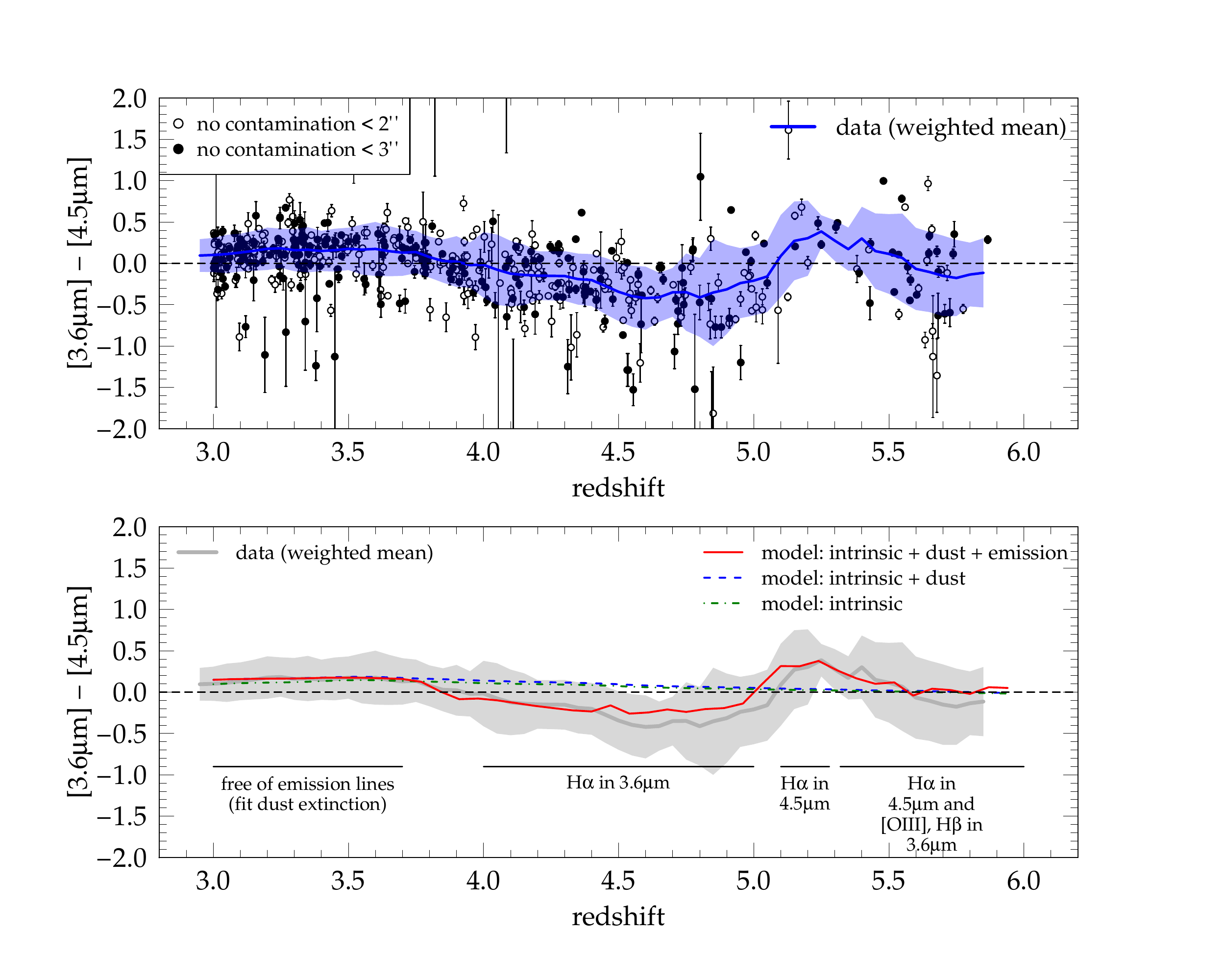}\\\vspace{-0.4cm}
\caption{\textbf{Top:} Observed color vs. redshift relation at $z>3$. The open (filled)  symbols denote galaxies with no contamination from companions within a radius of $2\arcsec$ ($3\arcsec$) in ACS/F18W and ground based data. The blue line shows the weighted mean relation with scatter (light blue band) for galaxies with no companion within $2\arcsec$.
\textbf{Bottom:} The best-fit intrinsic (blue, dashed), intrinsic $+$ dust (green, dot-dashed), and best-fit (red, solid) model.
\label{fig:fit1}}
\end{figure*}

	In \autoref{fig:intrinsiccolor}, we show the effect of intrinsic properties as well as dust attenuation on the observed \c3645~color on the example of a galaxy at $z=5$. First, we emphasize the small effect of metallicity on the continuum: a change from 1/5$^{{\rm th}}$ of solar to solar metallicity results in less than $0.05~{\rm mag}$ change in color.
	Second, in the case of an exponential increasing SFH with $\tau=500~{\rm Myr}$, the observed color as a function of age reddens less then $0.1~{\rm mag}$ for all possible ages of a $z=5$ galaxy. In the case of a constant and delayed SFH, the reddening is stronger due to domination by old stars with increasing age, but less than $0.2~{\rm mag}$ over a time of $\sim2~{\rm Gyrs}$, which corresponds to $z\sim 3$. Compared to this, the reddening by dust is of similar or larger amplitude (common dust extinctions are on the order of \ebmv$\sim 0.1-0.4~{\rm mag}$ in our sample).
	Note, that the model uncertainties are significantly reduced at high redshifts. First, the observed color is mostly independent of SFH for young galaxies with ages of less than $\sim1~{\rm Gyr}$, i.e., $z>5$. Second, it is expected that galaxies at high redshifts are dust poor \citep[e.g.,][]{DUNLOP13,BOUWENS14,CAPAK15}. Third, their age and metal content is well defined because of the young age of the universe at that time.
		
	Summarizing, we find that the expected reddening by dust exceeds the effect of metallicity as well as SFH and age for young galaxies up to $\sim1~{\rm Gyr}$ in age (corresponding to $z>5$). For galaxies with a constant or exponential declining SFH, as it is the case at lower redshifts, we expect the intrinsic color to change more significantly with age.  Also the color starts to increasingly depend on the assumed SFH for older galaxies and therefore lower redshifts.
	Finally, we note that a different IMF does not change these conclusions. For example, using a Salpeter IMF changes the observed color by less than $0.01~{\rm mag}$ at a given age.
	

	\subsubsection{Emission lines and dust}\label{sec:emlindust}
	
	Besides intrinsic properties, dust attenuation and emission lines contribute to the observed color of a galaxy.
	 We derive all rest-frame UV and optical emission lines relative to \halpha, which we vary in our model. In detail, we parametrize the evolution of the (rest-frame) \halpha~EW as
	
\begin{equation}\label{eq:ewha}
{\rm EW(H\alpha}) = {\rm EW(H\alpha)_0} \times \left( 1+z \right)^\alpha,
\end{equation}

where ${\rm EW(H\alpha)_0}$ and $\alpha$ are free fitting parameters.

Furthermore, we assume a constant (with redshift) line flux ratio \halpha/\oiii$=\xi$ within each of the three redshift windows (see \myrefsec{sec:zwindows}), because \oiii~only enters in a narrow redshift range at $z\sim2.2, 3.3, 5.5$ for windows (A), (B), and (C), respectively. In our case, \oiii~denotes the blended doublet and we assume \oiiired/\oiiiblue$\,=\,3$.

The \hbeta~line flux is determined from \halpha~assuming case B recombination,

\begin{equation}
{\rm f(H\alpha)/f(H\beta)} = 10^{0.4 \times  {\rm E(B-V)_{\rm neb}} \times (k_\beta - k_\alpha)  } \times 2.86,
\end{equation}

where $k_\beta$ and $k_\alpha$ are the coefficients for a given dust attenuation curve (we assume here \citet{CALZETTI00}\footnote{Several studies indicate that high redshift galaxies follow the dust attenuation curve similar to the one of the small Magellanic cloud. However, our model and data is not accurate enough to disentangle the effect of different attenuation curves.}) at the wavelengths of \halpha~and \hbeta, respectively.
The (stellar) dust extinction \ebmvstel \footnote{We assume \ebmvneb~=~\ebmvstel$/0.76$ \citep[][]{KASHINO13}. However, using a factor close to unity as suggested by more recent studies \citep{CULLEN14,SHIVAEI15,BARROS15} does not affect the results of this paper.} is parametrized as exponentially decreasing function of redshift \citep[e.g.,][]{HAYES11},

\begin{equation}\label{eq:dust}
{\rm E(B-V)_{{\rm stel}}} = {\rm E(B-V)}_{0} \times e^{-z/z_{d,0}}
\end{equation}

with $ {\rm E(B-V)}_{0}$ and $z_{d,0}$ as free parameters.
	We also model weaker optical emission lines (e.g., \SII, \NII, \hei), which we scale relative to the \hbeta~line fluxes according to \citet{ANDERS03}, assuming reasonable 1/5$^{{\rm th}}-1$ solar metallicity. Although not particularly strong in emission, these add up and can contribute up to $20\%$ to fluxes in the broad-band filters.
	
	The contributions of dust and emission lines are added to the intrinsic continuum described in the previous section. The emission lines are added assuming a full-width-at-half-maximum (FWHM) of $10~{\rm \AA}$ for the strong (\halpha, \hbeta, \oii, \oiii) and $5~{\rm \AA}$ for weak emission lines. Note that because of the large width of the broad-band filters, different (reasonable) choices of FWHM do not change the following results. 
	 The model colors are obtained by convolution of the generated SED with the corresponding filter transmission curves (Spitzer/IRAC for $3.6~{\rm \mu m}$ and $4.5~{\rm \mu m}$ and VISTA H and Ks band).

	\subsubsection{Fitting the observed color as a function of redshift}

	The top panel in \autoref{fig:fit1} shows the observed color in redshift window (C), i.e., $3 < z < 6$. The other two redshift windows are shown in \autoref{app2}. The symbols show individual galaxies (split in galaxies with no contamination within in $2\arcsec$ and $3\arcsec$ radius, respectively) and the blue band shows the running mean observed color (including $1\,\sigma$ scatter) as a function of redshift.
	In the following, we fit this observed color vs. redshift evolution with the previously described model for a given combination of metallicity and SFH.
	We use a Levenberg-Marquardt (LM) algorithm, as part of the \texttt{R/minpack.lm} package\footnote{\url{https://cran.r-project.org/web/packages/minpack.lm/index.html}} and proceed in two steps.
	
	\begin{enumerate}
	
	\item We fit the dust attenuation as a function of redshift (\autoref{eq:dust}) in regions devoid of emission lines (see \autoref{fig:emlins}). As mentioned above, this fit also includes intrinsic changes of the SED that are not taken into account elsewhere in our model. The sole purpose of this is to model the continuum below the optical emission lines (\hbeta, \oiii, \halpha).
	
	\item We fix the values of \ebmvz0 and $z_{{\rm d,0}}$ and fit the remaining parameters \EWhaz0, $\alpha$, and $\xi$ describing the emission lines.
	\end{enumerate}
	
	 This procedure is important to break the degeneracies between the effect of dust attenuation and emission lines on the observed color. We find that this is especially important at lower redshifts where galaxies show a significant amount of dust but much weaker EWs compared to high-z galaxies.
	
	We perform these two steps for in total four combinations describing our intrinsic SED: 1/5$^{{\rm th}}$ of solar and solar metallicity and two SFHs (constant and exponentially increasing\footnote{The delayed exponentially decreasing SFH yields similar results than a constant SFH and we do not list it here.}). The bottom panel in \autoref{fig:fit1} visualizes the fit for a constant SFH with solar metallicity at $3 < z < 6$. The best-fit model (intrinsic $+$ dust $+$ emission lines) in solid red is shown along with the dust reddened intrinsic color (blue dashed), and the intrinsic color (green dot-dashed).
	The horizontal lines label what affects the observed color in a given wavelength region (see also \autoref{fig:emlins}). The best-fit parameters for each redshift window and intrinsic SED are listed in \autoref{tab:fitresults}.


\begin{figure*}
\centering
\includegraphics[width=1.5\columnwidth, angle=0]{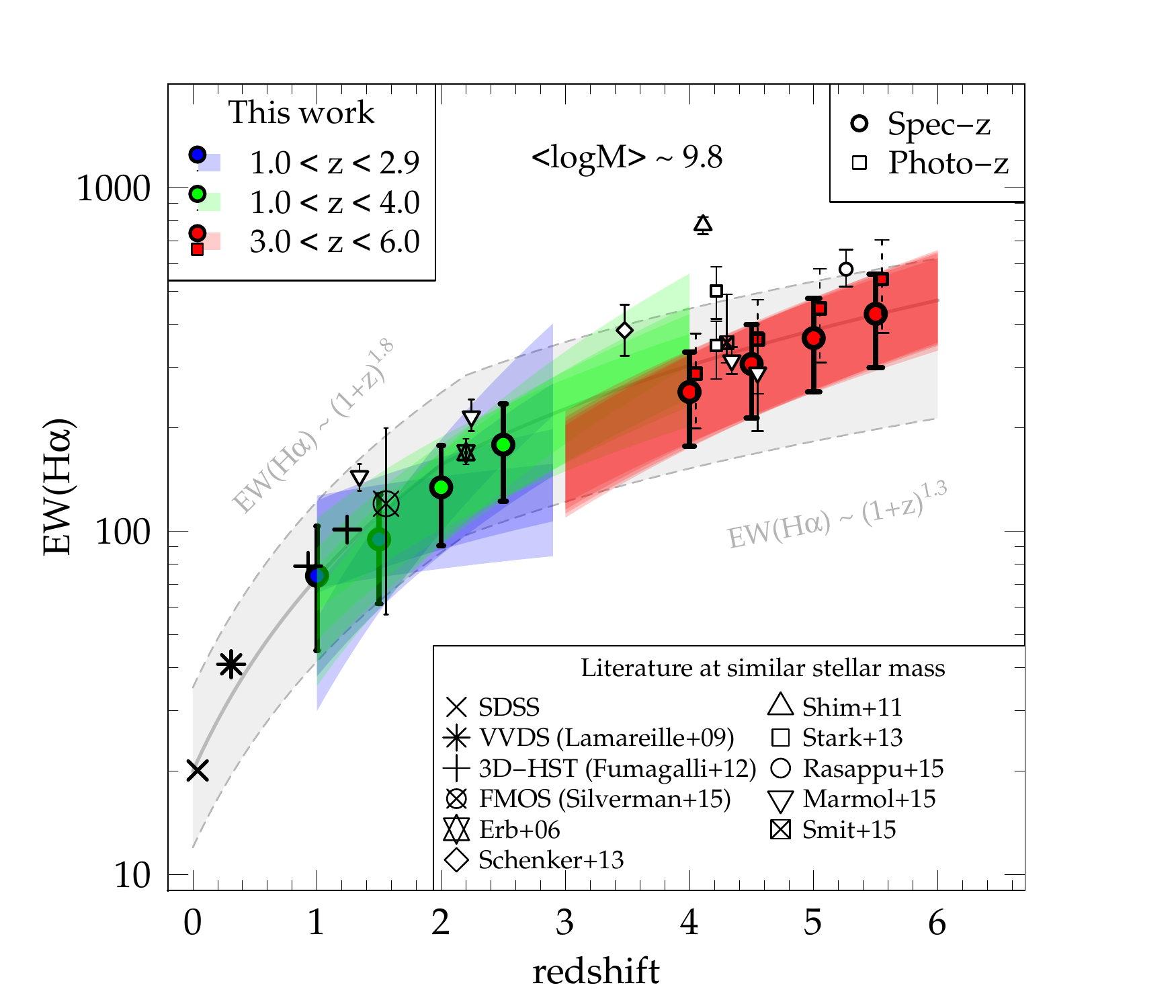}\\
\caption{Mean rest-frame \EWha~as a function of redshift. Our results in the three different redshift windows are shown in blue ($1.0 < z < 2.9$), green ($1.0 < z < 4.0$), and red ($3.0 < z < 6.0$). The different bands show the four combinations of metallicity and SFHs for each window (see text). The colored points show the redshifts where the \halpha~line is directly accessible. The red squares show the same for a sample based on photometric redshifts at $z>3$.
The symbols (see legend) show different studies measuring \EWha~directly from spectra \citep{ERB06,LAMAREILLE09,FUMAGALLI12,SILVERMAN15} or from observed color or SED fitting \citep{SHIM11,SCHENKER13,STARK13,RASAPPU15,MARMOL15,SMIT15a}.
To homogenize the results, we apply a constant factor of $15\%$ ($5\%$) to the literature measurements to correct for the \NII~and \SII~(\NII~only) emission lines where necessary.
\label{fig:results1}}
\end{figure*}

\section{Results}\label{sec:results}

The model described in the previous section allows us to fit the redshift dependence of the \halpha~EW as well as the \oiii/\halpha~line ratio from the observed color vs. redshift evolution. Furthermore, we are able to derive the sSFR$(z)$ from the former. The results are detailed in the following sections.

\begin{figure*}
\centering
\includegraphics[width=1.8\columnwidth, angle=0]{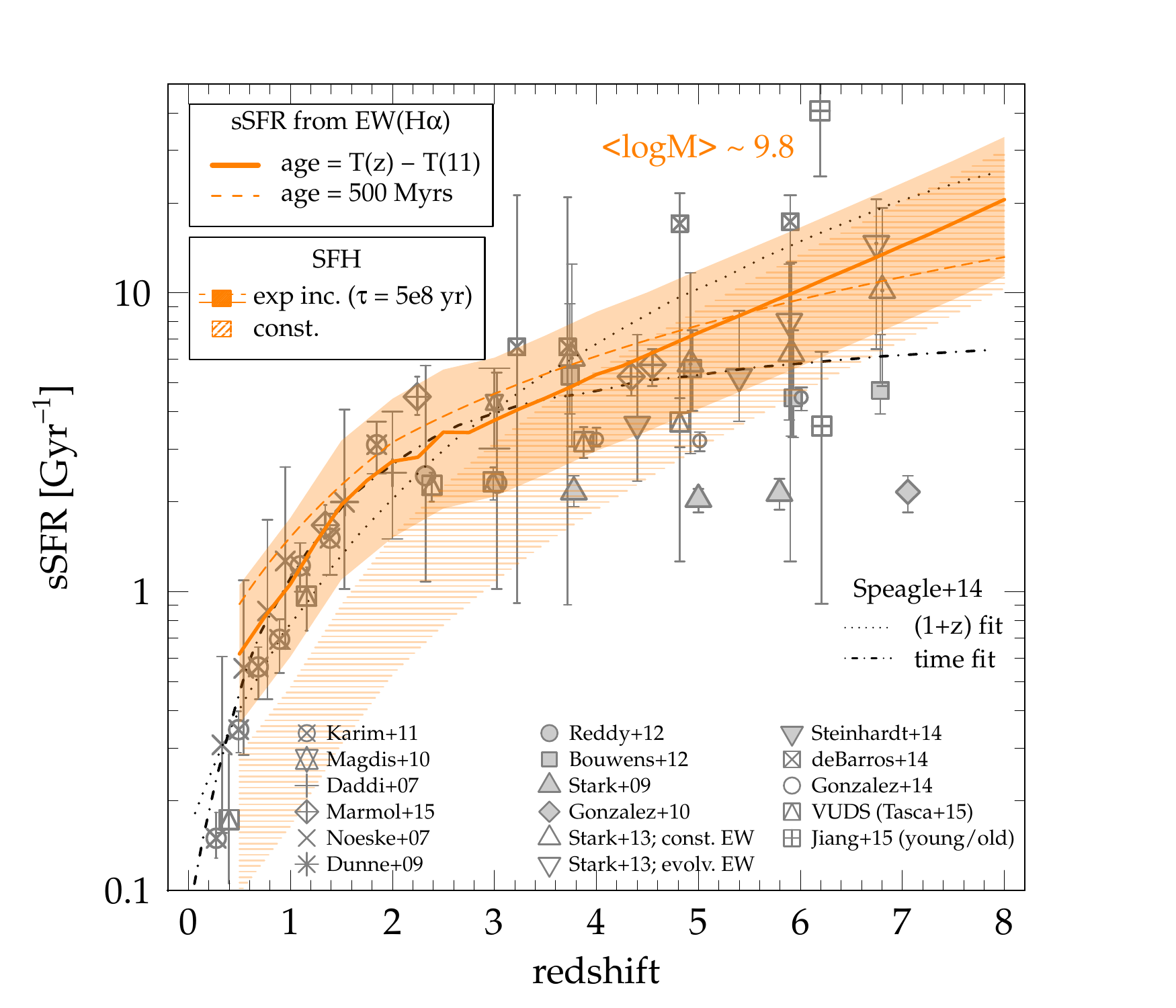}
\caption{Evolution of sSFR as a function of redshift computed from \EWha$(z)$ in the range $0.5 < z < 8.0$ (orange, extrapolated below $z=0.5$ and above $z=6$). The orange band assumes an exponentially increasing SFH ($\tau=5\times10^8~{\rm yr}$) and evolving age (Hubble time since $z=11$). The width of the band in includes a range in metallicity ($0.2-1.0~\Zsol$) and the range in \EWha. The orange dashed line shows the case for a fixes age of $500~{\rm Myrs}$. The hatched band shows the same for a constant SFH.
Along with this, we show observations at low-z \citep[][]{DADDI07,NOESKE07,DUNNE09,MAGDIS10,KARIM11,REDDY12}, and high-z without emission line correction \citep[][]{STARK09,GONZALEZ10,BOUWENS12} and with emission line correction \citep[][]{STARK13,GONZALEZ14,BARROS14,STEINHARDT14,TASCA15,JIANG15,MARMOL15}.
The dotted (dot-dashed) line shows the fit from \citet{SPEAGLE14} parametrized in redshift (time) space. We find the sSFR to be proportional to $(1+z)^{2.4}$ at $z<2.2$ and proportional to $(1+z)^{1.5}$ at higher redshifts, indicative of a flattening.
\label{fig:ssfr}}
\end{figure*}

\subsection{The \EWha~out to $z\sim6$}

\autoref{fig:results1} shows the redshift evolution of \EWha~for each of the three redshift windows at $1.0 < z < 2.9$, $1 < z < 4$, and $3 < z < 6$ (color-coded bands in blue, green, and red). We overlay the results from the four combinations of metallicity and SFH to show how our choice of intrinsic galaxy properties affects the results. As expected, the differences are negligible, verifying that the observed color is mainly driven by the contribution of emission lines (and dust), see \autoref{fig:intrinsiccolor}. The points color coded in the same way show \EWha~in the redshift ranges where it can be directly measured (see also \autoref{fig:emlins}). At $z>3$, we also show the results for a sample of galaxies selected by photometric redshifts (red squares). The EWs are consistent with our spectroscopic sample, suggesting that it is not severely biased towards young galaxies with enhanced star formation (see also \myrefsec{sec:biases}).

Our derived \EWha~in the redshift windows (A) and (B) are in excellent agreement with direct determinations from spectroscopy at $z<3$ obtained by \citet{ERB06} at $z\sim2$, \citet{FUMAGALLI12} \citep[at $1 < z < 2$ as part of 3D-HST;][]{VANDOKKUM11,BRAMMER12,SKELTON14}, and \citet{LAMAREILLE09} \citep[at $z\sim0.5$, as part of VVDS;][]{LEFEVRE05}.
Together with these studies based on spectroscopic measurements of the \halpha~emission line, our results agree with a strongly increasing \EWha~up to $z\sim2.5$, proportional to $(1+z)^{1.8}$ \citep[see also][]{FUMAGALLI12,SOBRAL14}.

This changes at higher redshifts, where we find that the \EWha~is evolving less steep than expected by the extrapolation from lower redshifts. This result is in good agreement with the recent study at $z\sim4.5$ \citep{MARMOL15,SMIT15a}, based on smaller samples but similar galaxy properties. The results of other studies \citep[][]{SHIM11,STARK13,SCHENKER13,RASAPPU15}\footnote{These studies do include \SII~and/or \NII~in their measurements of \halpha. We correct, if necessary, the contribution from \SII~and \NII~by assuming a constant factor of $15\%$, if both, and $5\%$, if only \NII, only.} show larger \EWha~on average, which we think is due to sample selection. On one hand, these galaxies are found in the faint tail of the luminosity distribution and stellar masses quoted for these galaxies are $0.5\,{\rm dex}$ or more lower than in our sample. On the other hand, in order to be spectroscopically confirmed, these continuum-faint galaxies have to be (strongly) \lya~emitting and therefore young with high star formation as it is expected that \EWha~is positively correlated with age and SFR \citep[e.g.,][]{LEITHERER99,COWIE11}.
Based on our minimally biased sample (see \myrefsec{sec:discussion}), we find that the evolution of the \halpha~EW is best parametrized by \EWha$\propto (1+z)^{1.3}$ at $z\gtrsim2.5$.

\begin{figure*}
\centering
\includegraphics[width=0.74\columnwidth, angle=0]{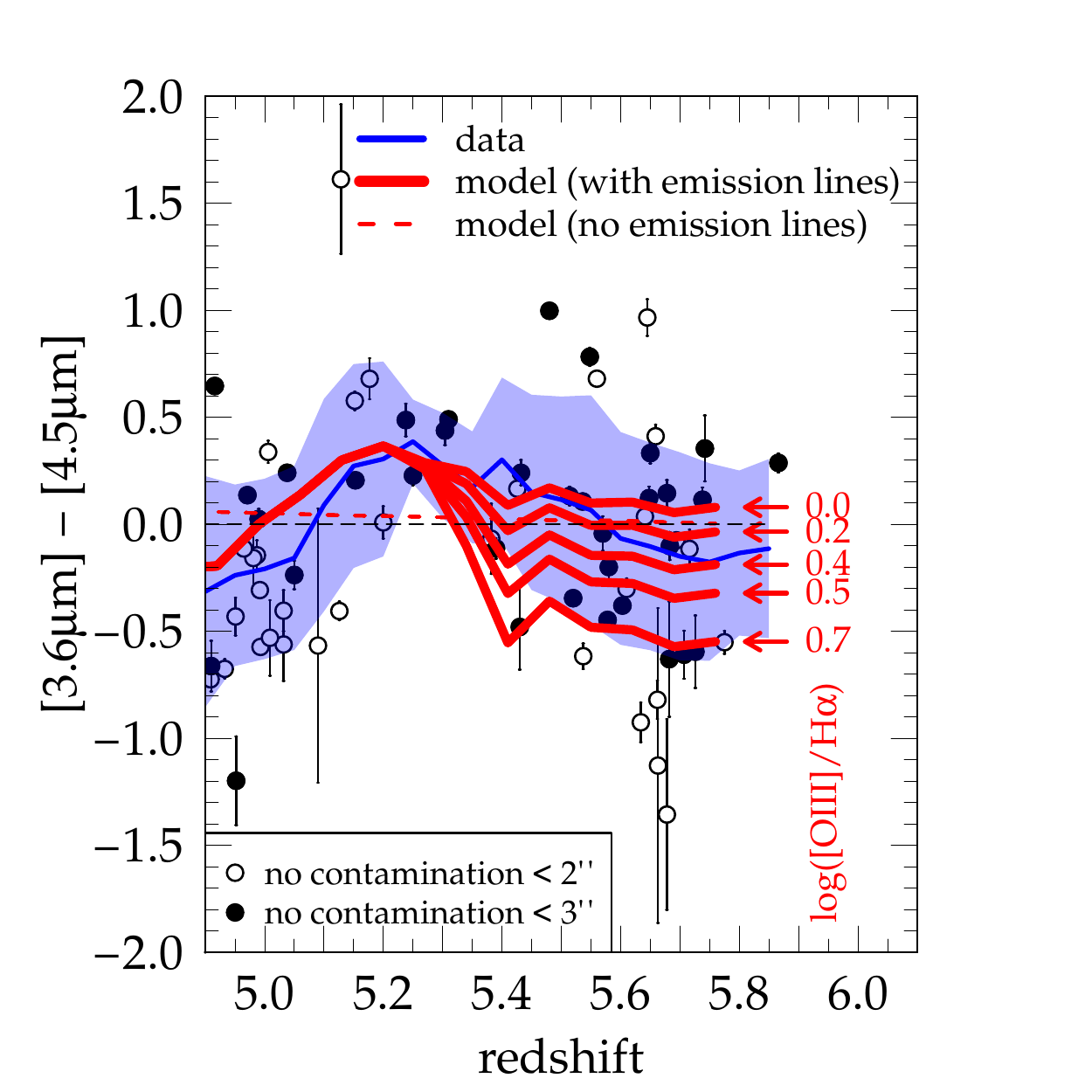}
\includegraphics[width=1.34\columnwidth, angle=0]{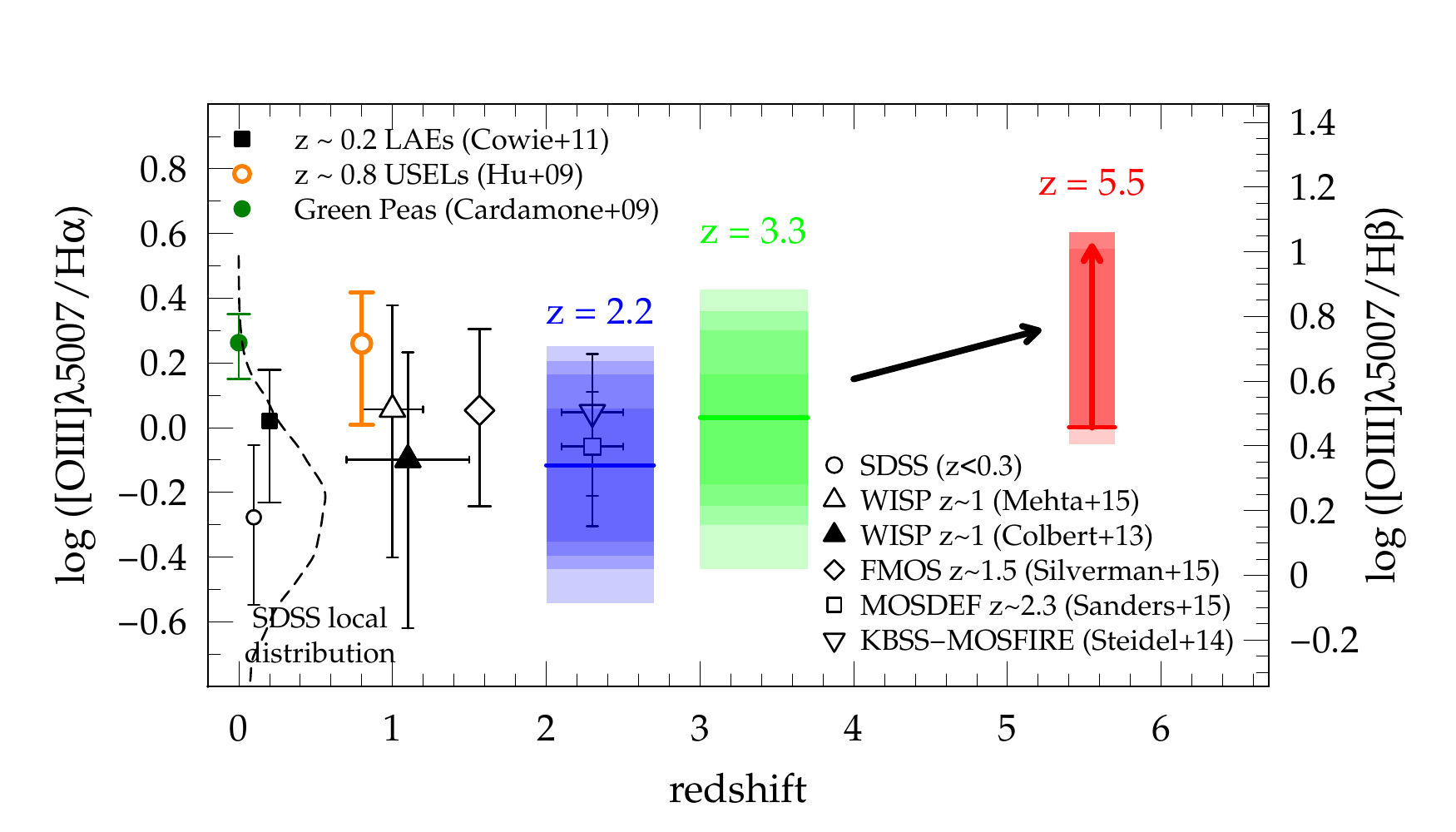}
\caption{\textbf{Left:} Because our data does not have the full redshift range to reliably measure the \oiii/\halpha~ratio at $z\sim5.5$, we show here the scatter of this measurement. The top red line with $\log($\oiii/\halpha$)=0.0$ shows the best-fit model. The other lines show models with increased \oiii/\halpha~ratios. We find $\log($\oiii/\halpha$)\sim0.0-0.7$ at $z\sim5.5$. The data is shown in blue and the model without emission lines is shown as red dashed line.
\textbf{Right:}  Mean dust corrected \oiii/\halpha~ratio as a function of redshift. All the samples are matched in stellar mass and the \oiii/\hbeta~ratio is computed assuming case B.
Our estimates based on broad-band colors are shown in blue ($z\sim2.2$), green ($z\sim3.3$), and red ($z\sim5.5$) including their uncertainties (from observation and different models). Spectroscopic measurement at lower redshifts are shown with symbols \citep{COWIE11,COLBERT13,STEIDEL14,MEHTA15,SANDERS15,SILVERMAN15} and the distribution of SDSS galaxies at $z\sim0$ is represented by the black dashed line (open circle: median). All in all, we find a progressively increasing \oiii/\halpha~ratio over the redshift range $z\sim2-6$. We also show high-z analogs as ``Green Peas'' \citep[green filled circle,][]{CARDAMONE09} and USELs at $z\sim0.8$ \citep[orange open circle,][]{HU09} for comparison.
\label{fig:results2}}
\end{figure*}

\subsection{The sSFR at $z>4$}
	
	The \halpha~is a tracer for star-formation and the stellar continuum red-ward of $4000~{\rm \AA}$ is a good tracer for stellar mass. Therefore, the \halpha~EW is directly proportional to the sSFR of a galaxy with the normalization factor solely depending on its internal properties such as metallicity, age of stellar populations, and SFH \citep[e.g.,][]{LEITHERER99,COWIE11}.
	The ensemble approach also allows a clear determination of the average and range of these properties and allows their propagation (forward-modeling) to the final results. This is one big advantage over a model based on ``galaxy-by-galaxy'' fitting. Furthermore, remember that our results are mostly insensitive to SFH, age, and metallicities at redshifts $z>4$ where the age of the universe is less than $1~{\rm Gyr}$ (see \autoref{fig:intrinsiccolor}).
		
	In order to convert the \EWha~to sSFR, we use \citet{BRUZUAL03} composite stellar population models bracketing different SFHs (exponentially increasing with $\tau~=~5~\times~10^8~{\rm yr}$ and constant). These models are stellar mass normalized and we can therefore directly convert \EWha~into a \textit{specific} \halpha~luminosity without any additional measurement of stellar mass. We then use the \citet{KENNICUTT12} relation (assuming a Chabrier IMF) to convert the specific \halpha~luminosity into a specific SFR. 
	\autoref{fig:ssfr} shows the resulting sSFR$(z)$ derived from our \EWha~evolution with redshift along with various measurement from the literature.
	The orange shaded band (and solid orange line) shows the sSFR derived based on the exponentially increasing SFH, while the hatched band shows the case of a constant SFH. For both we assume an age evolution corresponding to the cosmic time elapsed since $z=11$. The case for a constant age of $500~{\rm Myr}$ (and exponentially increasing SFH) is shown as dashed orange line.
	
	Since directly calculated from the \EWha, the sSFR evolution with redshift is different at low and high redshifts. While we find a strong increase of sSFR proportional to $(1+z)^{2.4}$ at $z\lesssim2.2$, this flattens out to a redshift dependence of $(1+z)^{1.5}$ at $z\gtrsim2.2$.
	
	For comparison, the symbols show various measurements from the literature, which are summarized in \autoref{tab:ssfrliterature1} (low redshift) and \autoref{tab:ssfrliterature2} (high redshift).
	These measurements can be broadly split into three groups:
	\textit{(i)} Measurements at $z\lesssim3$, which are based on reliable SFR indicators in the far-IR or sub-mm and without the problem of emission line contamination \citep{DADDI07,NOESKE07,DUNNE09,MAGDIS10,KARIM11,REDDY12},
	\textit{(ii)} measurements at $z>3$ that are based on SFR from UV and SED fitting as well as stellar mass estimates \textit{not} corrected for emission lines \citep{STARK09,GONZALEZ10,BOUWENS12}, and, 
	\textit{(iii)} measurements at $z>3$ that are based on SFR from UV and SED fitting as well as stellar mass estimates corrected for emission lines \citep{STARK13,GONZALEZ14,BARROS14,STEINHARDT14,JIANG15,TASCA15,MARMOL15}.
	



\subsection{The \oiii/\halpha~ratio at $z\sim6$}	\label{sec:oiiiha}

The ratio between \halpha~and \hbeta~solely depends on the dust attenuation.
Since we do fit \halpha~and dust attenuation, we can directly compute \hbeta~and thus separate it from the \oiii~line. This allows us to directly measure the ratio of \oiii~to \halpha~in three discrete redshift ranges centered on $z\sim2.2$, $z\sim3.3$ and $z\sim5.5$ (see \autoref{fig:emlins}).

	We are able to fit the \oiii/\halpha~ratio reliably at $z\sim2.2$ and $z\sim3.3$, since the redshift range at which the broad-band filters include \halpha~and \oiii~is fully covered by our data (see \autoref{fig:emlins} and \autoref{fig:fit1}). The right panel of \autoref{fig:results2} shows our \oiiired/\halpha~ratio\footnote{We split the \oiii~doublet ($4960\,{\rm \AA}$ and $5007\,{\rm \AA}$) by assuming \oiii$\lambda5007 = 1/3\,\times$~\oiii$\lambda4960$.} at $z\sim2.2$ and $z\sim3.3$ in blue and green, respectively, along with spectroscopically determined ratios \citep{STEIDEL14,SANDERS15}, which we find to be in excellent agreement with our measurements at $z\sim2.2$. The different bands for each color again show the four combinations of intrinsic properties in our model.
	
	At $z\sim5.5$ our data only includes galaxies up to $z\sim5.8$ and therefore does not cover the full redshift range (i.e., entry and exit of \oiii~in the $3.6\,{\rm \mu m}$ band) that is needed to reliably constrain the \oiii/\halpha~ratio. Furthermore, the sparse sampling of data at these redshifts contributes to the uncertainty. Also the addition of galaxies with photometric galaxies does not increase the sample size by much at $z>5.8$ as shown in \autoref{app}. We therefore discuss this case in more detail in the following.
	The left panel in \autoref{fig:results2} shows the redshift range from which we determine the \oiii/\halpha~ratio at $z\sim5.5$. As before, the points show the data and the blue band shows their scatter. The model without emission lines (but including dust) is shown as red dashed line. Note, that at these redshift, where the stellar ages are $<1\,{\rm Gyr}$, the model is very insensitive to SFH, age, and metallicity and primarily depends on the \oiii/\halpha~ratio in this case. We show the best fit \oiii/\halpha~ratio ($\log($\oiii/\halpha$)\sim0.0$) along with four different models with increasing ratios in red. This large variation over a range of $0.7~{\rm dex}$ in \oiii/\halpha~indicates the existence of galaxies with very strong \oiii~emission at these redshifts, in agreement with the recent findings at $z\sim6.7$ using a similar technique \citep[e.g.,][]{ROBERTSBORSANI15}. The galaxies clustering around $z\sim5.65$ are \lya narrow band selected and thus preferentially young and highly star forming. This could be the reason for their high \oiii/\halpha~ratios.
	The range of \oiii/\halpha~ratios at $z\sim5.5$ is shown in red on the right panel of \autoref{fig:results2}.

	We find a progressively increasing \oiiired/\halpha~ratio with redshift at $z>2$, once the large up-ward scatter at $z\sim6$ is taken into account. On the other hand, there is not much evolution between $z\sim1$ and $z\sim2$ \citep[literature at $z\sim1$ and $z\sim1.5$,][]{COLBERT13,MEHTA15,SILVERMAN15}. The average line ratio of local ($z\lesssim0.3$) SDSS galaxies is  $\sim0.2~{\rm dex}$ lower than at $z=2$ and $\sim0.2-0.8~{\rm dex}$ lower than at $z\sim6$. However, the distribution of \oiiired/\halpha~ratios in local galaxies (shown by the dashed density-histogram) is broad and sub-samples of these galaxies show similar line ratios as high-z galaxies. The potential of such ``local high-z analogs'' is further discussed in \myrefsec{sec:local}.


\section{Discussion}\label{sec:discussion}


We use a sample of $>500$ spectroscopically confirmed galaxies to derive the \halpha~EW, the sSFR$(z)$, and the \oiiired/\halpha~ratio at $3 < z < 6$. The main idea of our analysis is to base these measurements on primary observables (the observed color vs. redshift evolution) and to minimize the model uncertainties.
The forward modeling approach based on the ensemble instead of single galaxies allows us to marginalize over a range of SFH, metallicities, and ages and to propagate the uncertainties in these quantities to the final result. Due to the young age of the universe of less than $1~{\rm Gyr}$ at $z>4$, the differences between different assumptions that go into the modeling of the galaxy population are less significant, leading to robust results at these high redshifts (see also \autoref{fig:intrinsiccolor}).

In the previous section, we have established the following results.

\begin{enumerate}
\item The \EWha~increases continuously as $(1+z)^{1.8}$ up to $z\sim2.5$ and flattens off at higher redshifts with a redshift proportionality of $(1+z)^{1.3}$.

\item The sSFR increases proportional to $(1+z)^{2.4}$ at $z\lesssim2.2$ but shows a less strong evolution at higher redshifts proportional to $(1+z)^{1.5}$.

\item We find a best-fit \oiii/\halpha~ratio of $z\sim6$ star-forming galaxies on the order of unity (similar to $z=2$ and $z=3$ galaxies), however, with a scatter up to a ratio of five. This suggests the progressively increasing \oiii/\halpha~ratios at $z>3$.

\end{enumerate}

Before proceeding to the discussion of these results, we have to make sure that our sample is only minimally biased.

\subsection{How strong are the biases in our sample?}\label{sec:biases}

The emission line properties of galaxies vary substantially between different samples. This has a direct implication on the sSFR, since emission line strong galaxies tend to be young and strongly star forming. In particular, spectroscopic high-z samples are often emission-line and color selected and are therefore biased towards young, star forming galaxies.

The position of individual galaxies in our sample at $z>3$ on the stellar mass vs. SFR plane is in good agreement with the expected average star forming main-sequence extrapolated from lower redshifts measurements that are based on reliable IR SFR indicators \citep[e.g.,][]{SCHREIBER15,LEE15}. This is also verified by direct determination of the main-sequence at higher redshifts \citep[][]{STEINHARDT14}. However, this comparison has to be taken with a grain of salt as there are large uncertainties in the measurement of stellar masses and SFR from SED fitting for individual galaxies at high redshifts. We therefore proceed with two additional tests to quantify possible biases in our sample.

The VUDS sample is purely selected by photometric redshifts and is therefore less affected by the ``spectroscopic bias'' than the COSMOS sample, which is partially color selected. If the latter is severely biased, we would expect a non-negligible change in our results when removing the VUDS galaxies. However, this is not confirmed, which indicates that the biases are minimal.
We can even go one step further and select galaxies purely on their photometric redshift from the \textit{COSMOS2015} catalog. In this case, we do not expect any spectroscopic bias, on the other hand, these galaxies are obviously not spectroscopically confirmed and the photometric uncertainties tend to wash out the wiggles in the color vs. redshift relation. We therefore apply a strict selection on the photometric redshift errors as detailed in \autoref{app}. We perform the same analysis as for the spectroscopic sample and do not see significant differences in the results. As shown in \autoref{fig:results1} for $z>3$, the \halpha~EWs of the photometric sample are in good agreement with the determination from the spectroscopically selected samples.

From these various tests, we conclude that our total sample is minimally biased and represents well the average galaxy population at these redshifts and stellar masses. With this in hand, we continue to discuss the results obtained in the previous section.

\subsection{Broken sSFR evolution and the importance of major mergers}

The sSFR of the average star forming galaxy population is critical to understand galaxy formation in the high redshift universe. Different views of the evolution of sSFR at high redshifts exist not only between theoretical predictions and observations, but also amongst observations themselves.
	In particular, cosmological, hydrodynamical simulations predict a steep, continuous increase of sSFR over the whole redshift range up to very high redshifts. For example, an increase proportional to $(1+z)^{2.3}$ is expected in a picture where the galaxy growth is dominated by cold accretion \citep[e.g.,][]{DEKEL09}. Other hydrodynamical simulations, although under-estimating the sSFR at a given redshift compared to observations, are in favor of a continuously increasing sSFR up to high redshifts with a similarly steep redshift dependence \citep[e.g.,][]{DAVE11,SPARRE15}. Also, a strong redshift dependence with siilar exponent is expected if galaxy growth is closely connected to the dark matter halo assembly \citep[][]{TACCHELLA13}.
	Clearly, the sSFR of the average galaxy population has to increase towards higher redshifts in order to explain the findings of massive ($> 10^{11}\,\Msol$) galaxies found at high redshifts where they only have a couple of $100\,{\rm Myrs}$ to grow \citep[e.g.,][]{WEINMANN11,STEINHARDT14}. 
	Due to small sample sizes, biases, and uncertainties in the fitting of stellar masses and SFRs (mostly due to the unknown contribution of emission lines), it is not clear how strong the increase in sSFR actually is at $z>3$.
 
	Our results clearly show that the redshift evolution of the sSFR is broken. In particular we find a redshift proportionality of $(1+z)^{2.4}$ at $z<2.2$ and $(1+z)^{1.5}$ at higher redshifts. 
	At low redshifts ($z<2$), the steep increase in sSFR is in good agreement with the reliable measurements based on the far-IR, sub-mm, and spectroscopy. At very low redshifts, our model breaks down, resulting in a strong (factor three and more) over-estimation of sSFR. This because our method becomes increasingly more dependent on our assumptions on the SFH due to the older ages of the galaxies. This is indicated by the hatched band in \autoref{fig:ssfr} showing the case of a constant SFH, which is likely a better representation of the SFH of low-z galaxies compared to an exponentially increasing SFH. While these two SFH give similar results at $z>3$, they diverge substantially towards lower redshifts. Furthermore, with increasing age of the stellar population, the direct proportionality between \halpha~EW and sSFR is expected to break down.
	
	At $z\sim2.5$, the sSFR$(z)$ starts to flatten off and decreases the exponent of its redshift dependence from $2.4$ to $1.5$. However, the flattening is not as strong as suggested by other studies based on SED fitting, finding a $(1+z)$ exponent close to unity \citep[][]{GONZALEZ14,TASCA15,MARMOL15}. The average sSFR of $8-10\,{\rm Gyr}^{-1}$ at $z\sim5-6$ corresponds to an $e-$folding time for galaxy growth of $\sim100-200\,{\rm Myr}$, which is increased by a factor of two or more at higher redshifts. This is sufficient to explain the observations of galaxies with stellar masses of $\logm = 10.5-11.0$ at $z\sim5$.
	Even more massive galaxies ($\logm > 11.0$) have likely to be formed with the help of major mergers. This is in line with our finding that the redshift evolution of sSFR at $z>3$ is less steep than expected from simulation where galaxy growth is dominated by cold gas accretion \citep[e.g.,][]{DEKEL09}. We therefore do expect an additional mechanism for their mass growth. Major mergers -- important up to high redshifts \citep[e.g.,][]{TASCA14,RODRIGUEZGOMEZ15} -- can increase the galaxy stellar mass by factors of two per merger without increasing the sSFR over a long time scale\footnote{They might increase the sSFR on short time scales by triggering a star burst.}.


\begin{figure}
\centering
\includegraphics[width=1.15\columnwidth, angle=0]{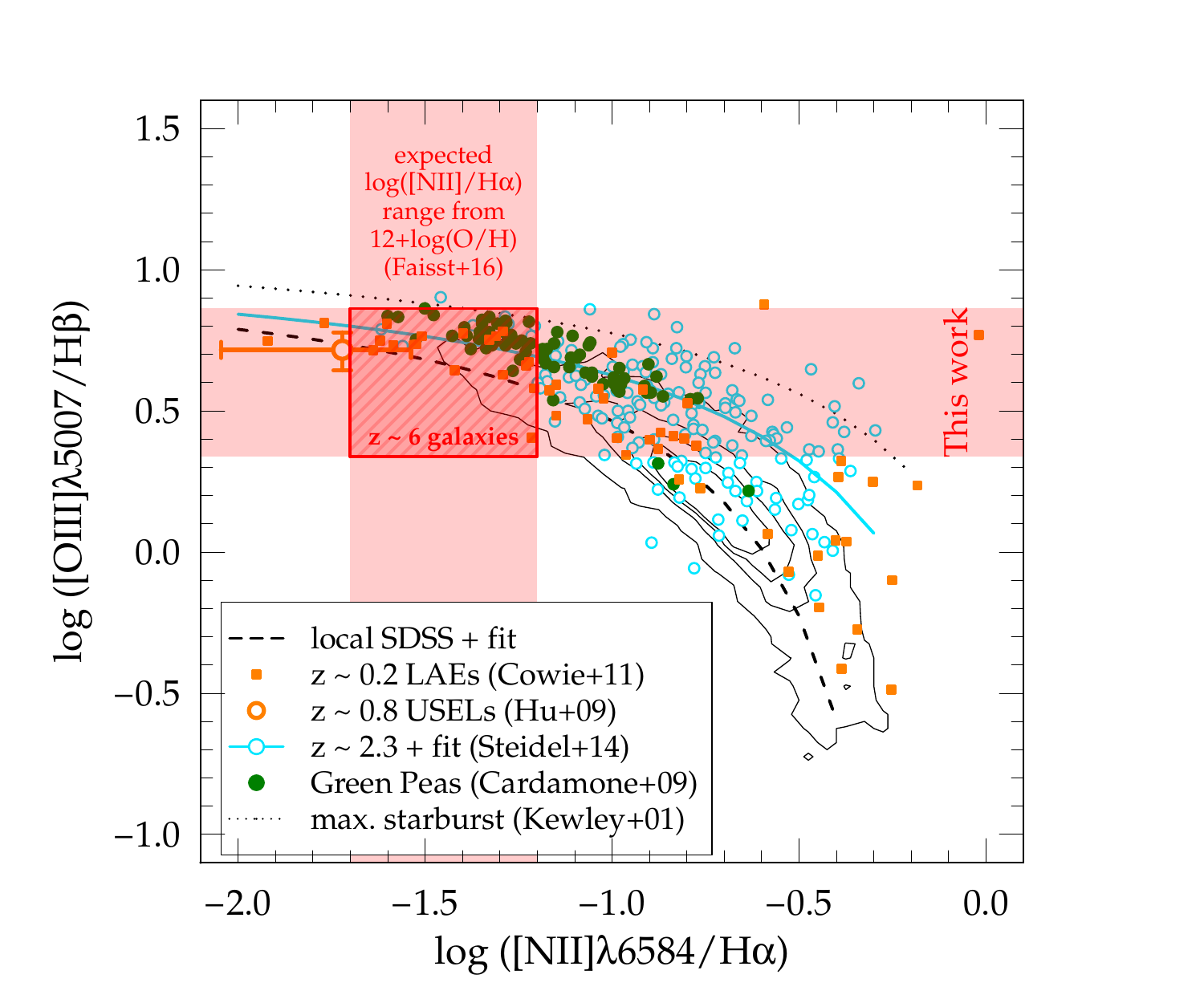}
\caption{The location of our $z\sim6$ galaxies (red-hatched square) on the BPT-diagram assuming the local \NII/\halpha~vs. metallicity relation \citep{MAIOLINO08} and average metallicities of $\oabund < 8.2$ at $z\sim6$ \citep{FAISST15}. Local SDSS galaxies are shown as contour with the best fit (dashed line). We also show LAEs at $z\sim0.2$ \citep[orange squares,][]{COWIE11}, USELs at $z\sim0.8$ \citep[orange point,][]{HU09}, $z\sim2$ galaxies \citep[cyan points and best fit as cyan line,][]{STEIDEL14} and ``Green Peas'' \citep[green dots,][]{CARDAMONE09}. Within our measurement uncertainties, we find the $z\sim6$ galaxies to lie in the expected region of the BPT-diagram, however, our data is not good enough to rule out a \NII~enhancement in high-z galaxies as currently discussed in the literature.
\label{fig:bpt}}
\end{figure}

\subsection{BPT-diagram at $z\sim6$}\label{sec:bpt}

The strong increase of sSFR shows that star-formation at early epochs paced at a different level than in today's galaxies. It is therefore a valid question, whether the scaling relations which are used in today's universe still hold for the very first galaxies.
	In particular, the \NII~abundance, a measure of metallicity, in young high-redshift galaxies is currently debated. It has been observed that intermediate redshift galaxies ($z\sim2$) reside in a different region on the ``Baldwin, Phillips \& Terlevich'' diagram \citep[BPT-diagram,][]{BALDWIN81,KEWLEY13} compared to the majority of local galaxies (see also \autoref{fig:bpt}). The BPT diagram connects the \oiiired/\hbeta~ratio with the \NII/\halpha~ratio, the latter being a tracer of the metal content in a galaxy \citep[e.g., ][]{PETTINI04,MAIOLINO08,KEWLEY08}.
	The offset between high and low redshift galaxies is not completely understood, yet, and studies argue for a harder stellar ionization field causing an enhanced \oiiired/\hbeta~ratio or a change in the electron temperature of high-z galaxies \citep[e.g.,][]{STEIDEL14}. Others are in favor of an enhancement of \NII~abundances in these galaxies with respect to local galaxies \citep[e.g.,][]{MASTERS14,SHAPLEY15,COWIE15}. Interestingly, also the stars in local globular clusters, which are thought to be formed in the very early universe, show nitrogen enhancements \citep[e.g.,][]{SPITE86,MACCARONE11}. If this is the case, a re-calibration of the local relation between \NII/\halpha~and metallicity at high redshifts is required.

	Although with a large uncertainty, we can test the above directly using our estimates of the \oiiired/\halpha~ratio at $z\sim6$. Including the scatter as described in \myrefsec{sec:oiiiha}, we find a range $\log($\oiiired/\hbeta$)= 0.35 - 0.85$, which we show on the $y$-axis of the BPT-diagram (\autoref{fig:bpt}). In addition, we expect the gas-phase metallicities of our galaxies to be on the order of $\oabund\sim8.2\pm0.2$ at $\logm\sim10.0$ \citep[][]{FAISST15}. Bluntly assuming that the \NII/\halpha~vs. metallicity relation holds at these redshifts, we would expect $\log($\NII/\halpha$)$ between -$1.7$ and -$1.2$ \citep{MAIOLINO08}, which is shown as a range on the $x$-axis on the BPT-diagram.
	
	As expected from their low metallicities, the location of our $z\sim6$ galaxies (shown by the red-hatched square) is located to the left of the locus of average local SDSS galaxies (black contours) as well as the bulk of $z\sim2$ galaxies \citep[cyan open points,][]{STEIDEL14}. On the other hand, they show a good agreement with the ``Green Peas'' (green dots), which are local galaxies with strong optical emission lines and are often referred to as local high-z analogs \citep{CARDAMONE09}. Similarly, the coincide with metal poor Ultra Strong Emission Line galaxies at $z\sim0.8$ \citep[USELS,][]{HU09,KAKAZU07} and metal poor strong \lya emitters at $z\sim0.2$ \citep[LAEs,][]{COWIE11}.

	In summary, the $z\sim6$ galaxies overlap with various high-z analogs at lower redshifts. Assuming similar physics in such galaxies as in high-z galaxies, this indicates that \NII~is a reasonable measure of metallicity at high-z. However, there is a lot of room to move and the uncertainties in our measurements are certainly too large to draw final conclusions and the idea of an enhancement of \NII~in high-z galaxies cannot be rejected.

\begin{figure}
\centering
\includegraphics[width=1.15\columnwidth, angle=0]{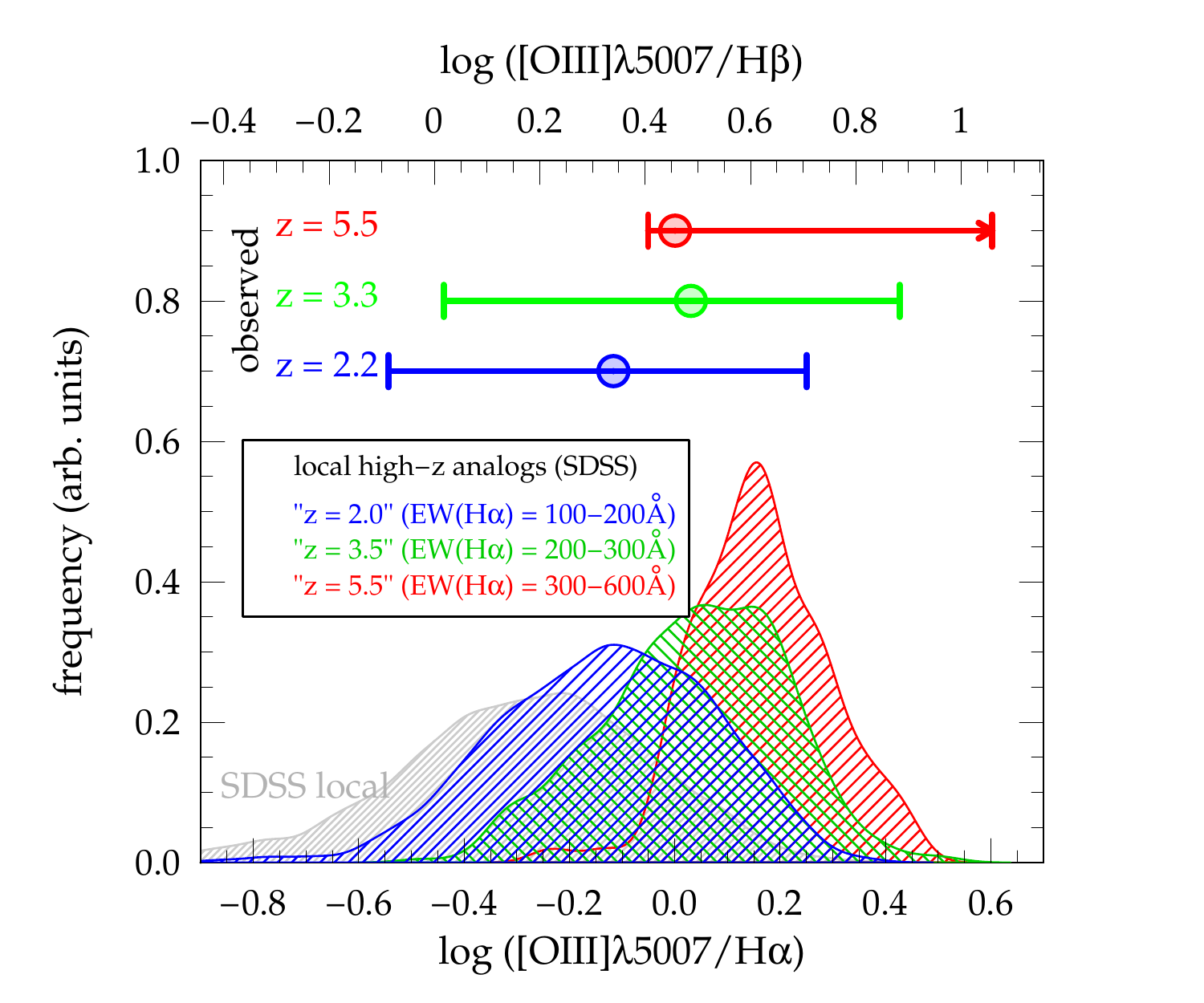}
\caption{High-z galaxies are ``not special''. Subsamples of local galaxies in SDSS that match the \oiiired/\halpha~ratios of high-z galaxies. The local high-z analogs are selected based on \EWha, i.e., sSFR, to be similar to $z~\sim~2$, $z~\sim~3.5$, and $z~\sim~5.5$ galaxies. This shows the potential of using local galaxies with high resolution spectra to investigate the properties of high-z galaxies.
\label{fig:local}}
\end{figure}

\subsection{Local high-z analogs}\label{sec:local}
	
	Clearly, as seen in \myrefsec{sec:bpt}, our efforts to measure the spectral properties of high-z galaxies are limited by the capabilities of the current (near-) IR telescopes.
	A method to progress is to define samples of local galaxies that resemble high-z galaxies in terms of spectral, photometric, and morphological properties (like ``Green Peas'' or USELs). These samples, providing high resolution spectral information, can be used to investigate the emission line properties of high-z galaxies (e.g., on the BPT-diagram) and provide useful priors on emission line strengths and ratios to improve the results from SED fitting.
	
	In \autoref{fig:local}, we show that SDSS galaxies selected by \textit{just} the \EWha~(i.e., sSFR) as they are expected at $z\sim2$, $z\sim3$ and $z\sim6$ have similar \oiiired/\halpha~distributions as high-z galaxies, i.e., could serve as high-z analogs. This indicates that high-z galaxies are not special in terms of their spectral properties, but they are included in sub-samples in the tails of the total distribution of low-z galaxies (shown in gray). This opens doors for investigating the spectral properties and physical relations of high-z galaxies using the high resolution spectra of galaxies in the local universe, and will be addressed in a future paper.

\section{Summary \& Conclusions}\label{sec:end}

In this paper, we introduce a simple method to measure the spectral properties and sSFRs of the \textit{ensemble} of high redshift galaxies via \textit{forward-modeling} of their observed color vs. redshift evolution.
	Our method does not require any SED fitting of stellar masses to measure the sSFR. The sSFRs are therefore derived as close as possible from \textit{primary observations} (the observed color of galaxies).
	Our ensemble approach allows a consistent modeling of the uncertainties of the various unknown intrinsic galaxy properties (metallicity, age, SFH) and to investigate their impact on the results. Importantly, it does not depend on the measurement of single galaxies, which is more uncertain.
 	This analysis is only made possible through the very large spectroscopic sample available on COSMOS, since accurate spectroscopic redshifts are necessary to get an accurate color vs. redshift relation.

Summarizing, these are our final conclusions.

\begin{enumerate}

\item We show that we are able to determine the spectral properties and subsequently the sSFR$(z)$ at $z>4$ from primary observations with very little uncertainties from modeling. This allows us to put important and reliable constrains on the physics of the first galaxies in our universe.


\item The sSFR increases proportional to $(1+z)^{2.4}$ at $z<2.2$ and proportional to $(1+z)^{1.5}$ at higher redshifts. This indicates a fast build-up of stellar mass in galaxies at $z>3$ within $e-$folding times of $100-200\,{\rm Myrs}$. The redshift evolution at $z~>~2.2$ cannot be explained by cold accretion driven growth alone.

\item We find a tentative increase in the \oiiired/\hbeta~ratio between $z\sim2$ and $z\sim6$, however, this has to be confirmed by larger samples at $z>5$.

\item Taking a face value the \oiiired/\hbeta~ratio and assuming the \NII/\halpha~vs. metallicity relation of local galaxies, we find $z\sim6$ galaxies to reside at a similar location on the BPT-diagram as the ``Green Peas'' as well as metal poor USELs and LAEs. Our data does not allow us to draw further conclusions on a possible \NII~enhancement in high-z galaxies as it is currently being debated.

\item High-z analogs can be selected from the tail distribution of local SDSS galaxies by matching in sSFR. Local galaxies are therefore a powerful tool to investigate the spectral (and physical) properties of high-z galaxies and also provide useful priors on emission line strengths that can be used to improve the SED fitting. 

\end{enumerate}

The spectral properties of high-z galaxies will ultimately be tested by the IR capability of the next generations of telescopes, most importantly the JWST, launched in the next years. Our sample provides a useful test-bed.


\acknowledgments

We would like to thank Josh Speagle, Micaela Bagley, and Bahram Mobasher for valuable discussions. AF acknowledges support from the Swiss National Science Foundation.
JDS is supported by JSPS KAKENHI Grant Number 26400221, the World Premier International Research Center Initiative (WPI), MEXT, Japan and by CREST, JST.
Parts of this work is based on data obtained with the European Southern Observatory Very Large Telescope, Paranal, Chile, under Large Programs 175.A-0839 and 185.A-0791. This work is supported by funding from the European Research Council Advanced Grant ERC-2010-AdG-268107-EARLY.
Based on data products from observations made with ESO Telescopes at the La Silla Paranal Observatory under ESO programme ID 179.A-2005 and on data products produced by TERAPIX and the Cambridge Astronomy Survey Unit on behalf of the UltraVISTA consortium.
Also, the authors wish to recognize and acknowledge the very significant cultural role and reverence that the summit of Mauna Kea has always had within the indigenous Hawaiian community. We are most fortunate to have the opportunity to conduct observations from this mountain.

\bibliographystyle{apj}
\bibliography{../bibli.bib}

\appendix
\section{Comparison to photo-z sample of galaxies at $z>3$}\label{app}

Spectroscopically selected galaxy samples at high redshift ($z\gtrsim4$) could be biased towards young, star-forming galaxies with strong \lya~emission that would increase \EWha~and sSFR compared to the average population. However, due to the target selection of our spectroscopic samples (especially the VUDS sample, which selection is based on photometric redshifts) we do not expect severe biases. We can (at least partly) assess the severeness of biases at $3 < z < 6$ by comparing our sample to photometrically selected galaxies.
	
	For the investigation of the observed color vs. redshift relation, we need a clean sample of photometrically selected galaxies. The basis of our sample builds on the \textit{COSMOS2015} photometric catalog, containing photometric redshifts that are derived with more than $30$ filters including broad-, intermediate-, and narrow-bands. The photometric redshifts are verified with large numbers of spectroscopically confirmed galaxies and show an accuracy of better than $\sim5\%$ on average \citep[see][]{LAIGLE15}.
	We select a clean sample of galaxies by requiring $68\%$ of the probability distribution function within $3 < z < 6$ and a redshift uncertainty less than $5\%$. The former rejects galaxies with a considerable second redshift solution at $z<3$ and the latter results in $\Delta~z \leq 0.27$ in this redshift range, which is enough to resolve the ``wiggles'' in the color vs. redshift relation caused by emission lines. We reject galaxies which have companions within $2\arcsec$ as in the case of the spectroscopic sample. Finally, we adjust the stellar mass range of the photometric sample to be similar as the spectroscopic sample at $z>3$ ($\left< \logm \right> \sim 9.8$).
	
\begin{figure*}
\centering
\includegraphics[width=0.8\columnwidth, angle=0]{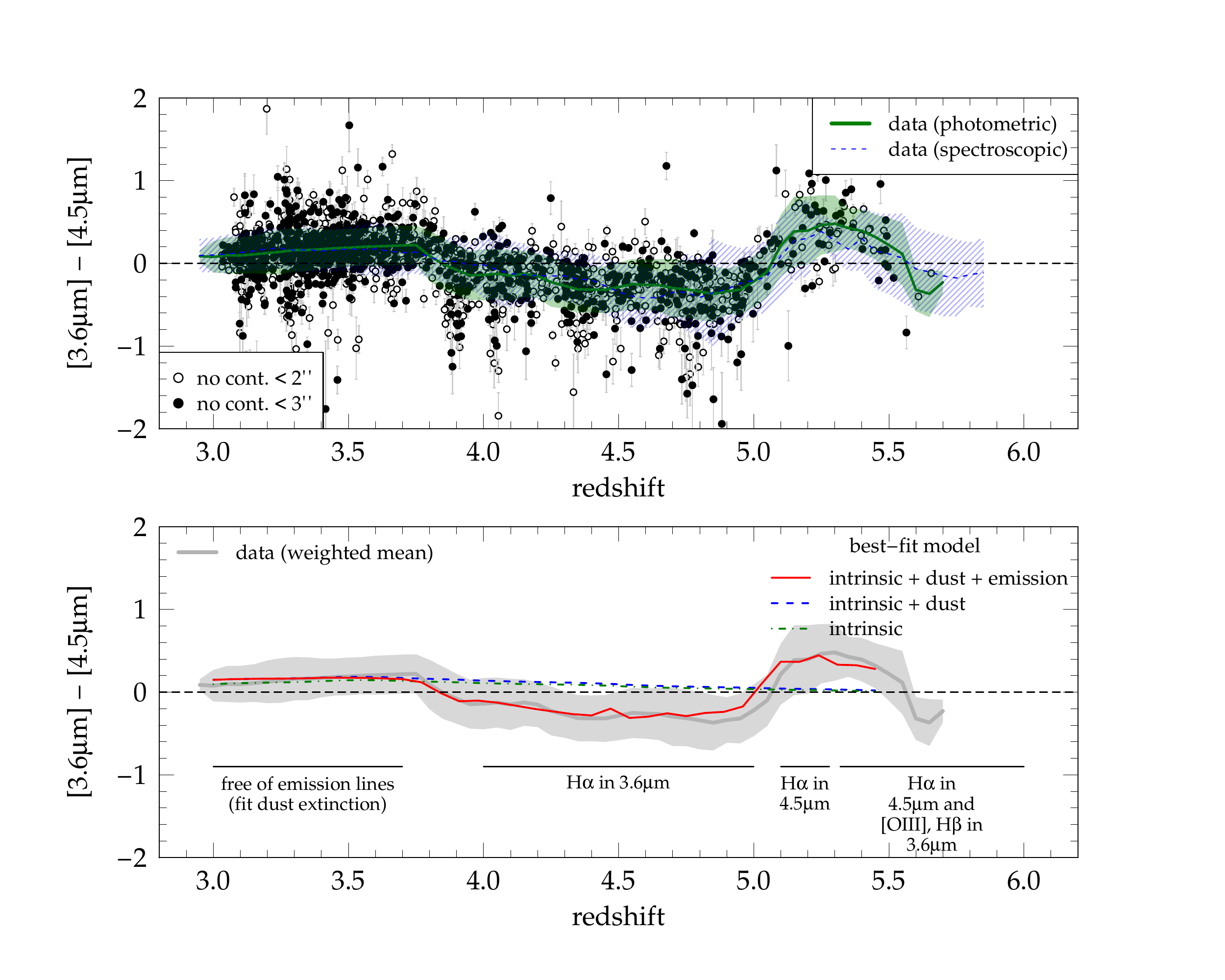}\\
\caption{Same as \autoref{fig:fit1} but for galaxies selected by their photometric redshift at $3 < z_{{\rm phot}} < 6$ (see text). The blue hatched region shows the median color from the spectroscopic sample. The green band shows the median color from the photometric sample, which is in agreement with the spectroscopic one within $1\sigma$. This shows that there are no significant biases in our spectroscopic sample.
\label{fig:fit_photz}}
\end{figure*}

	The top panel in \autoref{fig:fit_photz} shows the observed color vs. redshift evolution for our photometric galaxy sample in the range $3 < z_{{\rm phot}} < 6$ in green. The weighted mean of the spectroscopic sample is shown in blue. We already see that the distributions are very similar.
	As for the spectroscopic sample, we fit our multi-component model to the observed color. The example of the best-fit model for a constant SFH with solar metallicity is shown on the bottom panel of \autoref{fig:fit_photz} in red. The resulting \EWha$(z)$ is in good agreement with the one obtained from the spectroscopic sample (see \autoref{fig:results1}).
	All in all, we conclude that our sample is minimally biased and represents well the average population of galaxies at $z>3$.

	

\clearpage
\newpage
\section{Observed color vs. redshift evolution at $1 < z < 4$}\label{app2}

The redshift windows (A) ($1.0 < z < 2.9$) and (B) ($1 < z < 4$) are used to verify our method. The results are compared to \EWha~measurements directly based on spectroscopic observations. In the following, we show the same figures as \autoref{fig:fit1} for these redshift windows.

\begin{figure}[b]
\centering
\includegraphics[width=0.7\columnwidth, angle=0]{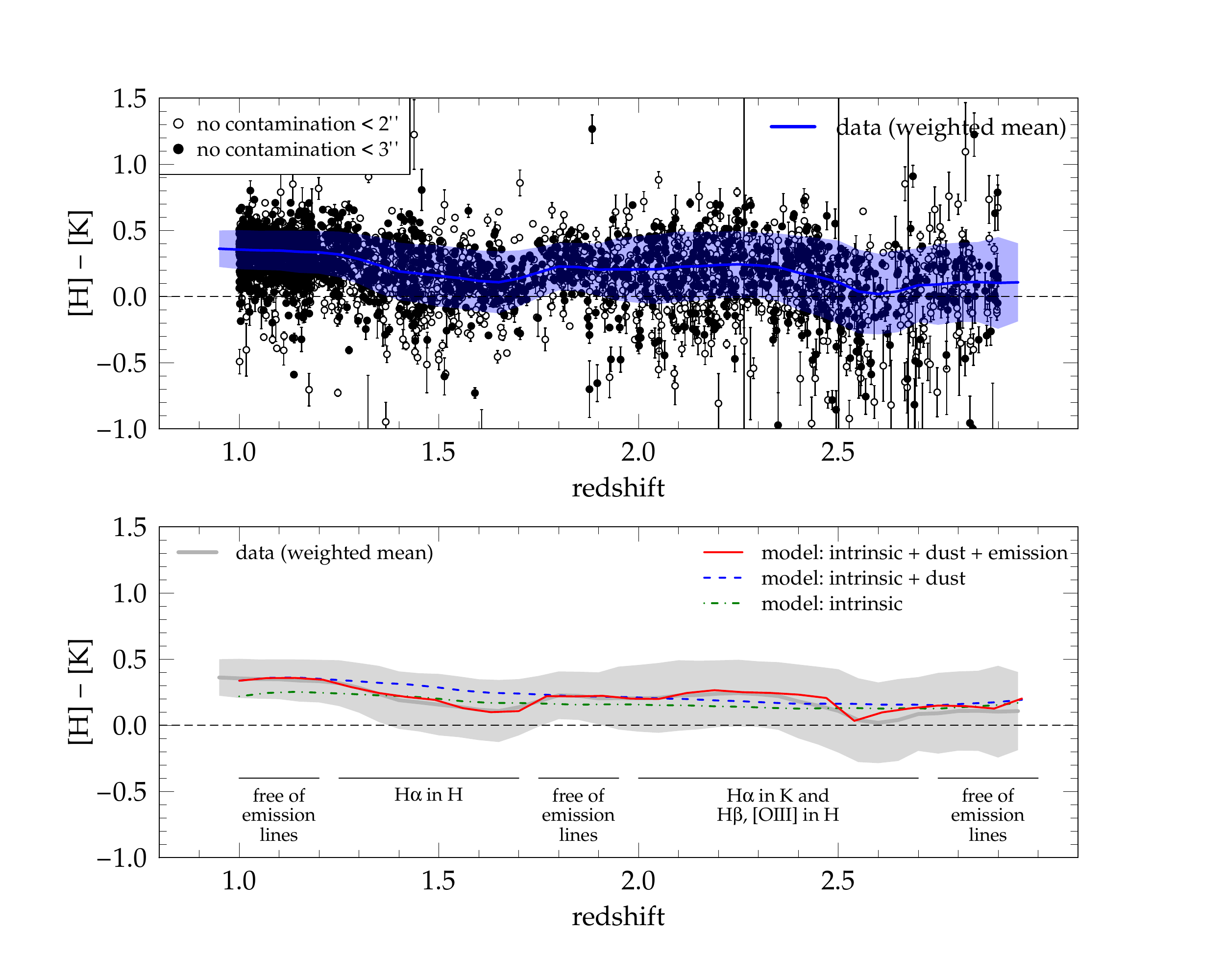}\\
\includegraphics[width=0.7\columnwidth, angle=0]{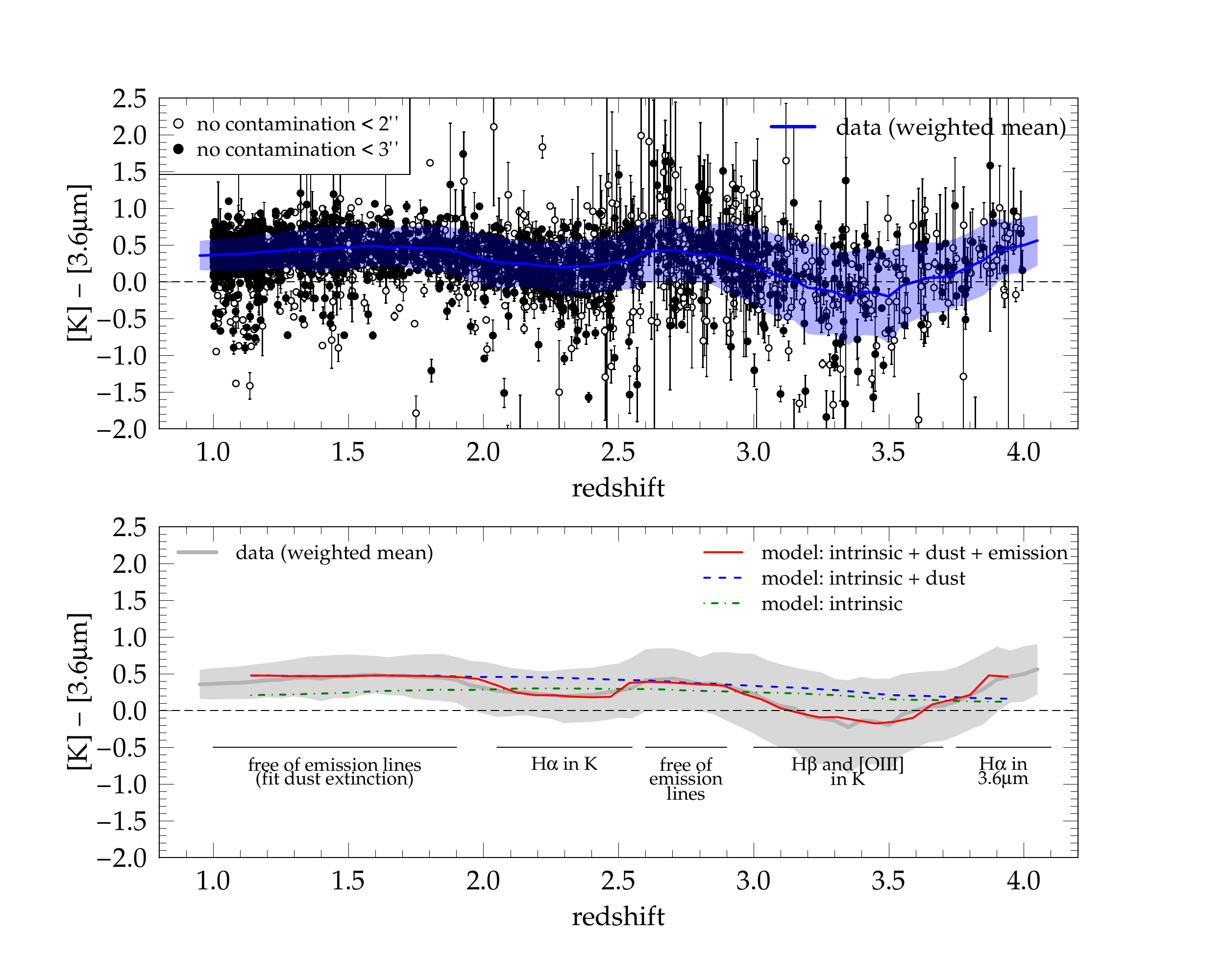}\\
\vspace{-0.4cm}
\caption{Same as \autoref{fig:fit1} but for redshift windows (A) ($1.0 < z < 2.9$) and (B) ($1 < z < 4$). \textbf{Top panels:} Observed color vs. redshift relation. The open (filled)  symbols denote galaxies with no contamination from companions within a radius of $2\arcsec$ ($3\arcsec$) in ACS/F18W and ground based data. The blue line shows the weighted mean relation with scatter (light blue band) for galaxies with no companion within $2\arcsec$.
\textbf{Bottom panels:} The best-fit intrinsic (blue, dashed), intrinsic $+$ dust (green, dot-dashed), and best-fit (red, solid) model.
\label{fig:fit_app}}
\end{figure}

\clearpage
\begin{landscape}
\capstartfalse
\begin{deluxetable}{p{1.9cm} p{1.3cm} p{3.2cm} p{2.8cm} p{3.5cm} p{4cm} p{4cm}}
\tabletypesize{\scriptsize}
\tablecaption{Literature on sSFR measurements at low redshift\label{tab:ssfrliterature1}}
\tablewidth{0pt}
\tablehead{
\multicolumn{2}{c}{Sample} & \multicolumn{3}{c}{Measurements} & \multicolumn{1}{c}{~} & \multicolumn{1}{c}{~}\\
\multicolumn{2}{c}{---------------------------------------} & \multicolumn{3}{c}{---------------------------------------------------------------------------------------} & \multicolumn{1}{c}{~} & \multicolumn{1}{c}{~}\\
\colhead{$z$} & \colhead{\# galaxies} & \colhead{SFR [M$_{\odot}$/yr]} & \colhead{M [M$_{\odot}$]} & \colhead{Emission lines} & \colhead{Comments} & \colhead{Reference}
}

\startdata\\[0.1cm]

$1.4 < z < 2.5$ & 1300$^{*,\dagger}$ & different estimators (UV, far-IR, $1.4~{\rm GHz}$) & SED fitting & none & $-$ & \citet{DADDI07} \\[0.5cm]

$0.2 < z < 1.1$ & $\sim2900$$^{\dagger}$ & optical emission lines and $24~{\rm \mu m}$ & SED fitting & none & $-$ & \citet{NOESKE07} \\[0.5cm]

$0.2 < z < 3$ & $\sim 23.000$$^*$ & stacked $1.4~{\rm GHz}$ & SED fitting & none & $-$ & \citet{DUNNE09} \\[0.5cm]

$z\sim3$ & 248$^{\dagger}$ & UV & SED fitting & none & $-$ & \citet{MAGDIS10} \\[0.5cm]

$0.2 < z < 3$ & $>10^5$$^*$ & stacked $1.4~{\rm GHz}$ & SED fitting including $24~{\rm \mu m}$ & none & $-$ & \citet{KARIM11} \\[0.5cm]

$1.4 < z < 3.7$ & $\sim300$$^{\dagger}$ & SED, UV, and $24~{\rm \mu m}$ & SED fitting & none & $-$ & \citet{REDDY12} 

\enddata
\tablenotetext{$\dagger$}{spectroscopic sample}
\tablenotetext{$*$}{photometric sample}

\end{deluxetable}
\capstarttrue
\clearpage
\end{landscape}

\clearpage
\begin{landscape}
\capstartfalse
\begin{deluxetable}{p{1.9cm} p{1.3cm} p{3.2cm} p{2.8cm} p{3.5cm} p{4cm} p{4cm}}
\tabletypesize{\scriptsize}
\tablecaption{Literature on sSFR measurements at high redshift\label{tab:ssfrliterature2}}
\tablewidth{0pt}
\tablehead{
\multicolumn{2}{c}{Sample} & \multicolumn{3}{c}{Measurements} & \multicolumn{1}{c}{~} & \multicolumn{1}{c}{~}\\
\multicolumn{2}{c}{---------------------------------------} & \multicolumn{3}{c}{---------------------------------------------------------------------------------------} & \multicolumn{1}{c}{~} & \multicolumn{1}{c}{~}\\
\colhead{$z$} & \colhead{\# galaxies} & \colhead{SFR [M$_{\odot}$/yr]} & \colhead{M [M$_{\odot}$]} & \colhead{Emission lines} & \colhead{Comments} & \colhead{Reference}
}

\startdata\\[0.0cm]

\multicolumn{7}{c}{\textbf{This work}}\\[0.1cm]
\hline\\

$1 < z < 6$ & $\sim3600$$^{\dagger}$ & (specific) \halpha~luminosity & $-$ & corrected, direct measurement from observed color & Not involving fitting of stellar mass or SFR from UV. Strongly increasing sSFR at $z<2.5$ and flattening off at higher redshifts & \textbf{This work} \\[0cm]

\\[0.1cm]
\multicolumn{7}{c}{\textbf{Studies \textit{not} including emission lines}}\\[0.2cm]
\hline\\

$4 < z < 6$ & $\sim800$$^{*}$ & UV (not dust corrected) & SED fitting & none & flat sSFR$(z)$ relation, not considering emission lines & \citet{STARK09} \\[0cm]

$z\sim7$ & 11$^*$ & UV & SED fitting & none & flat sSFR$(z)$ relation, not considering emission lines  & \citet{GONZALEZ10} \\[0.5cm]

$4 < z < 7$ & $\sim2400$$^*$ & UV & UV mass to light ratios & none & flat sSFR$(z)$ relation, not considering emission lines & \citet{BOUWENS12} \\[0.5cm]

\\[0.1cm]
\multicolumn{7}{c}{\textbf{Studies including emission lines}}\\[0.2cm]
\hline\\

$3.8 < z < 5$ & 92$^{\dagger}$ & UV (dust and emission line corrected) & SED fitting & included, obtained from observed color & nebular emission added to SED templates from observed \EWha~distribution. Extrapolated to $z\sim7$ by assuming constant \EWha~as well as \EWha$\propto (1+z)^{1.8}$. Find strongly increasing sSFR at $z>4$ & \citet{STARK13} \\[0.5cm]

$4 < z < 6$ & $\sim750$$^*$ & from SFH derived by SED fitting & SED fitting & included, assuming constant \EWha~and $\propto(1+z)^{1.52}$ from $z\sim2$. Emission line contribution subtracted from photometry before fitting. & increasing sSFR$(z)$ for increasing \EWha~model & \citet{GONZALEZ14} \\[0.5cm]

$3 < z < 6$ & $\sim1700$$^{*}$ & from SFH derived by SED fitting & SED fitting & included, proportional to Lyman continuum photon production & strongly increasing sSFR at $z>4$ & \citet{BARROS14} \\[0.5cm]

$0.1 < z < 5$ & $\sim4500$$^{\dagger}$ & SED fitting & SED fitting & included, proportional to UV photons  & shallow evolution of sSFR$(z)$ at $z>3$ & \citet{TASCA15} \\[0.5cm]

$z\sim6$ & 27$^{\dagger}$ & UV and \lya & SED fitting & included, from \lya & based on bright LBGs and LAEs. Find two populations (split at ages of $30~{\rm Myr}$) with vastly different sSFR & \citet{JIANG15} \\[0.5cm]

$1.2 < z < 5$ & $\sim400$$^{\dagger,*}$ & \halpha~luminosity & SED fitting including emission line templates. Use SED SFR to fix \halpha~flux and other emission lines related to it. & residual from SED fitting (excluding contaminated filters) & shallow evolution of sSFR at $z>3$. & \citet{MARMOL15} 

\enddata
\tablenotetext{$\dagger$}{spectroscopic sample}
\tablenotetext{$*$}{photometric sample}

\end{deluxetable}
\capstarttrue
\clearpage
\end{landscape}

\end{document}